\documentclass[showpacs,preprintnumbers,amsmath,amssymb,twocolumn,nofootinbib,nobibnotes,longbibliography]{revtex4-1}
\usepackage{color}
\usepackage[colorlinks,urlcolor=blue]{hyperref}
\usepackage{epstopdf}
\usepackage{graphicx}
\usepackage[normalem]{ulem}

\usepackage[english]{babel}
\usepackage{tabularx}

\usepackage{dcolumn}
\usepackage{multirow}
\usepackage{bm}
\usepackage{upgreek}

\begin{document}

\title{
Valley Hall effect caused by the phonon and photon drag
}

\author{M.~M.~Glazov}
\author{L.~E.~Golub} 	
\affiliation{Ioffe Institute,  	194021 St.~Petersburg, Russia}


\begin{abstract}
Valley Hall effect is an appearance of the valley current in the direction transverse to the electric current. We develop the microscopic theory of the valley Hall effect in two-dimensional semiconductors where the electrons are dragged by the phonons or photons. We derive and analyze all relevant contributions to the valley current including the skew-scattering effects together with the anomalous contributions caused by the side-jumps and the anomalous velocity. The partial compensation of the anomalous contributions is studied in detail. The role of two-phonon and two-impurity scattering processes is analyzed. We also compare the valley Hall effect under the drag conditions and the valley Hall effect caused by the external static electric field.
\end{abstract}

\maketitle

\section{Introduction}

The Hall effect is the generation of the electric current transversal to both the external electric and magnetic fields. The anomalous Hall effects are the phenomena where the transversal flux generation is not directly related to the action of the Lorentz force on the particles~\cite{Hall:1881aa,RevModPhys.82.1539,dyakonov_book}. The prominent examples of the anomalous Hall effects are  the Hall effect in magnetic metals and semiconductors, the spin Hall effect resulting in generation of the spin current transversal to the electric one, and the valley Hall effect (VHE) in multivalley systems. In the latter situation the particles in different valleys of the Brillouin zone propagate in the opposite directions.
VHE is now in focus of theoretical and experimental investigations~\cite{Xiao:2012cr,Mak27062014,PhysRevB.90.075430,Jin893,Onga:2017aa,Unuchek:2019aa,Lundt:2019aa,PhysRevLett.122.256801,urbig} owing to the emergence of a novel material system, transition metal dichalcogenide monolayers (TMDC MLs). In this strictly two-dimensional semiconductors the electrons reside in the two time-reversal related valleys $\bm K_\pm$ at the Brillouin zone edges, and chiral selection rules at optical transitions provide a direct optical access to the valley physics.

The anomalous and valley Hall currents can be generated without an external electric field provided a directed electron flux was created by other means. An important example is the electron drag by the non-equilibrium phonons in the system (phonon drag effect) or due to 
an alternating electromagnetic fields (photon drag effect)~\cite{PhysRevLett.122.256801}. 
This type of experiments has the advantage of studying the Hall fluxes of not only charged but also neutral particles. 
A prominent example is the excitonic VHE in TMD MLs which, owing to recent progress in accessing the exciton transport, appears in the focus of modern research~\cite{PhysRevLett.120.207401,Lundt:2019aa,Unuchek:2019aa}.

\begin{figure*}[t!]
 \includegraphics[width=0.8\linewidth]{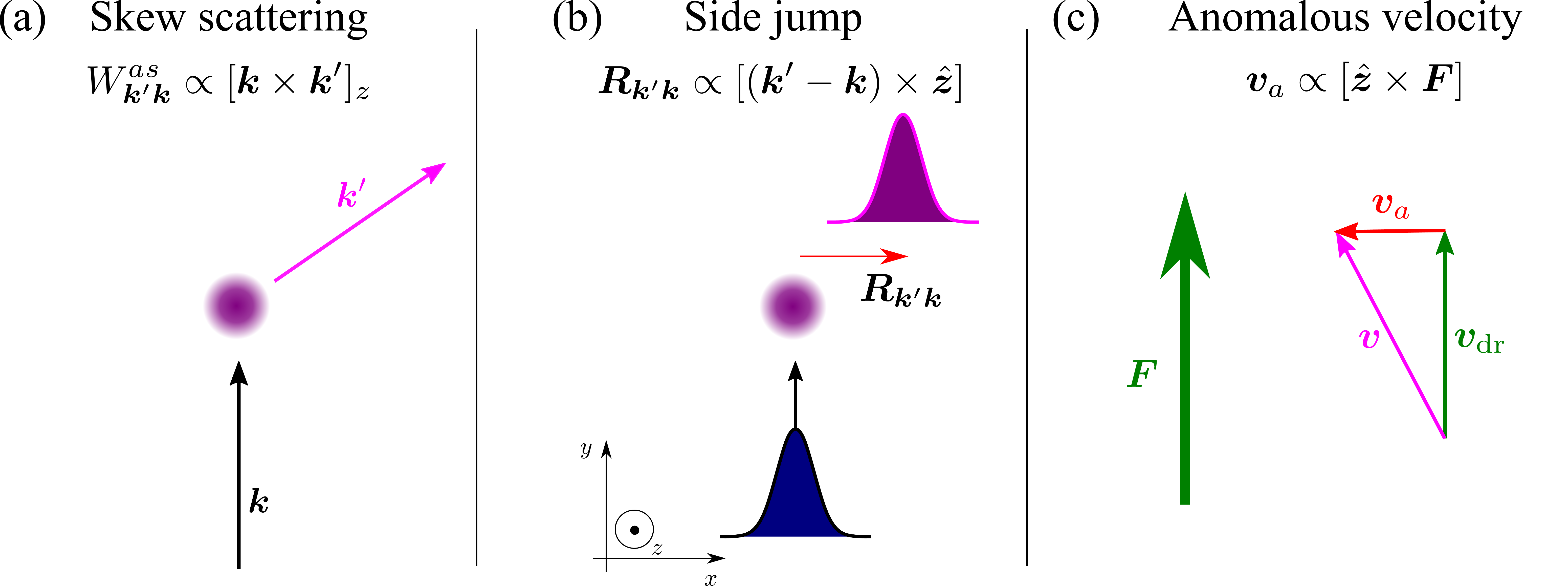}
\caption{\textbf{Mechanisms of the valley Hall effect.} (a) Skew scattering, Sec.~\ref{sec:skew}; (b) side jump and (c) anomalous velocity, Sec.~\ref{sec:side}. Electrons are pushed along $y$-axis by the force $\bm F$ and transversal component of the current appears. The effects are shown only for one valley, $\bm K_+$, in the $\bm K_-$ valley the direction of the transversal motion is opposite. Here $\bm k$ and $\bm k'$ are the initial and final wavevectors in the scattering process, $\bm v_{\rm dr}$ is the drift velocity, $\bm v_a$ is the anomalous velocity. Inset in (b) demonstrates the coordinate frame. 
}\label{fig:intro}
\end{figure*}

Despite the fact that a theory of anomalous Hall effect has a long history, 
a complete theory of VHE under the drag conditions is still missing.
It is firmly established that the spin-orbit interaction is at the origin of the anomalous Hall effect~\cite{PhysRev.95.1154,SMIT1955877,SMIT195839,Sinitsyn_2007,Ado_2015,PhysRevLett.123.126603,rev_Niu}. Three specific mechanisms illustrated in Fig.~\ref{fig:intro}, namely, (a)~the asymmetric or skew-scattering, (b) the side-jump, and (c) the anomalous velocity or Berry phase have been identified to give rise to the anomalous Hall effect in nonmagnetic systems~\cite{dyakonov_book}. Furthermore, it turns out that there are substantial cancellations of the anomalous velocity and side-jump contributions in semiconductors with parabolic bands~\cite{dyakonov_book,PhysRevB.75.045315}. 
However, by now, 
VHE
has been described mainly in the terms of the anomalous velocity~\cite{rev_Niu,Xiao:2012cr,Onga:2017aa,PhysRevLett.122.256801,Gianfrate:2020aa}, while the role of the skew-scattering and side-jump contributions has not been addressed in detail.
Moreover,  
the drag-induced VHE
requires a special analysis, because it is not immediately obvious whether anomalous velocity contributions could play a role in the absence of a static electric field.

Our paper aims to provide a consistent theory of the electron valley Hall effect under the drag conditions.
We present the microscopic theory of the electron VHE 
where the electron flux is initially created by the directed flow of the phonons or the photons.
We take into account all the contributions to the effect, including the skew-scattering and the side-jump along with the anomalous velocity terms.  
Also, for completeness and illustration of particular contributions, we provide the calculation of the valley Hall effect caused by a static electric field in two-dimensional (2D) Dirac materials.
The paper is organized as follows: Sec.~\ref{sec:model} outlines the band structure model used in our calculations. Further we analyze the contributions to the valley Hall effect due to the asymmetric scattering of electrons, Sec.~\ref{sec:skew}, and due to the side-jump and anomalous velocity, Sec.~\ref{sec:side}. Section~\ref{sec:discuss} contains the discussion of our results with the emphasis on the cancellation effects of the side-jump and anomalous velocity contributions. Brief conclusion and outlook are presented in Sec.~\ref{sec:concl}.

\section{Model}\label{sec:model}

We develop the theory of the electron VHE in 2D materials 
within the minimal two-band model, where the electron Hamiltonian describing the states in the vicinity of the $\bm K_+$ point of the Brillouin zone has the form~\cite{2053-1583-2-2-022001,Xiao:2012cr}
\begin{equation}
\label{H:el}
\mathcal H_+ = \begin{bmatrix}
0 &\gamma(k_x - \mathrm i k_y)\\
\gamma(k_x + \mathrm i k_y) & - E_g
\end{bmatrix}, \quad \gamma = \frac{\hbar p_{cv}}{m_0} 
{\in \mathbb R.}
\end{equation}
Here $E_g$ is the bandgap, $p_{cv}$ is the interband momentum matrix element, $k_\pm = k_x \pm \mathrm ik_y$, $x$ and $y$ are the axes in the 2D plane. 
The Hamiltonian describing the states in the vicinity of the $\bm K_-$ point can be obtained from Eq.~\eqref{H:el} by the time-reversal symmetry, yielding $k_\pm \to -k_\mp$. We note that the Hamiltonian~\eqref{H:el} can be used to describe both spin-up and spin-down pairs of the conduction and valence band states with proper choice of $E_g$, where the appropriate combinations of the conduction and valence band spin-orbit coupling constants are included.

While the Hamiltonian~\eqref{H:el} admits analytical diagonalization which makes it possible to calculate the VHE for electrons in electric fields at the arbitrary ratio of the electron kinetic energy $\varepsilon_k$ to the band gap $E_g$~\cite{PhysRevB.75.045315,Ado_2015,PhysRevB.96.235148} we present the results in the lowest non-vanishing order in $\varepsilon_k/E_g \ll 1$. The reasons are as follows: The model Hamiltonian becomes inadequate for $\varepsilon_k/E_g {\sim 1}$ where the non-parabolic corrections to the electron dispersion due to other bands disregarded in Eq.~\eqref{H:el} become important. Also, the topological properties of the Hamiltonian~\eqref{H:el} are ill-defined (the Chern numbers are $\pm 1/2$ in contrast to the general theory predicting integer Chern numbers for the Bloch bands); at the same time, the anomalous velocity contributions are sensitive to the topological properties of the Hamiltonian. We address this issue in Sec.~\ref{sec:discuss}, where the extensions of the model are briefly discussed. 

Accordingly, we take the electron dispersion in the form $\varepsilon_k = \hbar^2 k^2/2m$, where $m =\hbar^2 E_g/(2\gamma^2)$  is the effective mass, and present the electron Bloch function in the conduction band at the $\bm K_+$ valley
\begin{subequations}
\label{Blochs}
\begin{equation}
\label{e:Bloch}
u_{c,\bm k}(\bm r) \approx  c_{\bm k}\left(|c\rangle + \frac{\gamma}{E_g} k_+ |v\rangle \right), \quad c_{\bm k} = 1-\frac{\gamma^2}{2{E_g^2}} k^2,
\end{equation}
where $|c\rangle =(1,0)^{\rm T}$ and $|v\rangle = (0,1)^{\rm T}$. The analogous expression holds for the valence band Bloch function:
\begin{equation}
\label{v:Bloch}
u_{v,\bm k}(\bm r) \approx  c_{\bm k}\left(|v\rangle - \frac{\gamma}{E_g} k_- |{c}\rangle \right), \quad c_{\bm k} = 1-\frac{\gamma^2}{2{E_g^2}} k^2.
\end{equation}
\end{subequations}
 At the $\bm K_-$ valley the replacement $k_\pm \to -k_\mp$ in Eqs.~\eqref{Blochs} is needed.

Let the scattering potential be
\begin{equation}
\label{V:scatt}
{\mathcal V}(\bm r) = \begin{bmatrix}
 V_c(\bm r) & 0\\
 0 & V_v(\bm r)
 \end{bmatrix},
\end{equation}
where $V_c(\bm r)$ and $V_v(\bm r)$ are the potential energies related to a static disorder and acoustic phonons. 
As a model of the static disorder we consider the short-range defects and present the potential in the form
\begin{equation}
\label{v:dis}
V_{c,v}(\bm r) = \sum_{i} U_{c,v} \delta(\bm r - \bm R_i),
\end{equation}
where $\bm R_i$ are the random positions of the defects, and $U_{c,v}$ are the parameters. For the electron-phonon interaction we consider the deformation potential scattering by longitudinal acoustic phonons and recast the potential energies in the form~\cite{gantmakher87,PhysRevB.85.115317}
\begin{equation}
\label{v:def}
V_{c,v}(\bm r) = \sum_{\bm q} \mathrm i \Xi_{c,v} \sqrt{\frac{\hbar q}{2\rho s }}  \left(b_{\bm q} e^{\mathrm i \bm q \bm r - \mathrm i sqt} -  {\rm H.c.}\right),
\end{equation}
where $\Xi_c$ and $\Xi_v$ are deformation potentials for the conduction and valence bands, $\rho$ is the two-dimensional mass density of the 2D system, 
$s$ is the speed of sound, $\bm q$ is the phonon in-plane wavevector, ${b}_{\bm q}$ (${b}_{\bm q}^\dag$) are the phonon annihilation (creation) operators, and the normalization area is set to unity. To describe the skew scattering effect by acoustic phonons we need to go beyond Eq.~\eqref{v:def} and consider two-phonon processes~\cite{gy61,bs78}, which are specified below in Sec.~\ref{sec:skew}.
Generally, $V_c \ne V_v$, for example, at the electron-acoustic phonon scattering the conduction and valence band deformation potentials are strongly different in TMDC MLs~\cite{PhysRevB.90.045422,Phuc:2018aa}.

Using Eq.~\eqref{e:Bloch} we calculate the matrix element for the scattering where both initial ${(\bm k)}$ and final ${(\bm k')}$ states are in the conduction band:
\begin{equation}
\label{Mcc}
M_{\bm k',\bm k} = V_c(q) + \mathrm i  \xi V_v(q) [\bm k' \times \bm k]_z, \quad \xi= \frac{\gamma^2}{E_g^2}.
\end{equation}
Here,  the non-parabolicity-induced valley-independent corrections $\sim k^2$ are disregarded.
We also derive the interband scattering matrix elements as
\begin{subequations}
\label{Mcvvc}
\begin{align}
&M^{cv}_{\bm k',\bm k} = \frac{\gamma}{E_g}[V_v(q) k'_- - V_c(q) k_-], \\
&M^{vc}_{\bm k',\bm k} = \frac{\gamma}{E_g}[V_v(q) k_+ - V_c(q) k_+'] \equiv (M^{cv}_{\bm k,\bm k'})^*.
\end{align}
\end{subequations}
Here $V_{c,v}(q)$ are the Fourier components of the potentials $V_{c,v}(\bm r)$ in Eq.~\eqref{V:scatt}. Note that for the considered static disorder model, Eq.~\eqref{v:dis}, the quantities $V_{c}(q)$ and $V_{v}(q)$ are simply given by $U_c$ and $U_v$, respectively, and do not depend on the transferred wavevector. Similar situation holds for the electron-phonon scattering: In what follows we consider only quasi-elastic processes of the electron-phonon scattering at sufficiently high temperatures ($k_B T \gg ms^2$ for non-degenerate electrons). 
Thus, electron scattering by low energy phonons, $\hbar s q \ll k_B T$, for both the phonon emission and absorption is equivalent to the short-range defect scattering with
\begin{equation}
\label{Vcvq:ph}
{n_i |U_{c,v}|^2 = \Xi_{c,v}^2\frac{k_B T}{2\rho s^2 }}
\end{equation}
where $n_i$ is the impurity density.

Equations~\eqref{Mcc} and \eqref{Mcvvc} are presented for the $\bm K_+$ valley. To obtain the matrix elements for the electrons in the $\bm K_-$ valley the replacements $\xi \to -\xi$ and $k_\pm \leftrightarrow -k_\mp$ are needed. 

For completeness we also present the expression for the Berry curvature {in the conduction band} at $\bm k=\bm 0$ at the $\bm K_+$ valley:
\begin{equation}
\label{Bcurv}
\bm{\mathcal F} = 2\Im{ \left\langle\left.\frac{\partial u_{{c},\bm k}}{\partial k_y} \right| \frac{\partial u_{{c},\bm k}}{\partial k_x}\right\rangle }\hat{\bm z}
=  -2\xi \hat{\bm z},
\end{equation}
with $\hat{\bm z}$ being the unit vector along the normal to the 2D plane.  The anomalous velocity in the presence of the real force field $\bm F$ (which enters the Hamiltonian in the form of $-\bm F\cdot \bm r$) acting on the electron takes the form
\begin{equation}
\label{v:anom}
\bm v_a = {1\over \hbar} \bm{\mathcal F} \times \bm F.
\end{equation}
The Berry curvature and the anomalous velocity have opposite signs for the electrons at the $\bm K_-$ valley.

In what follows we are interested in the VHE, that is generation of the opposite electric currents in $\bm K_+$ and $\bm K_-$ valleys. Accordingly, we define the valley current as
\begin{equation}
\label{VHC:def}
\bm j_\text{VH}=\frac{1}{2}(\bm j_{\bm K_+}-\bm j_{\bm K_-}),
\end{equation}
where $\bm j_{\bm K_\pm}$ are the electric currents in the corresponding valleys. It follows from the symmetry analysis that $\bm j_{\rm VH}\perp \bm F$, where $\bm F$ is the external force or the dragging force acting on the electrons. Thus, below we calculate the transversal to the force component of the current in the $\bm K_+$ valley only. Since the corresponding component in the $\bm K_-$ valley is opposite in the direction and has the same absolute value, our calculation gives the valley Hall current. 

While we focus here on the VHE, the same theory applies for the spin Hall effect in the systems where the Hamiltonians $\mathcal H_\pm$, Eq.~\eqref{H:el}, describe spin branches, e.g., in 2D semiconductors and topological insulators.

\section{VHE due to skew scattering}\label{sec:skew}

It is convenient to describe the electron valley transport in the framework of the kinetic equation approach. We consider the valley $\bm K_+$ and introduce $f_{\bm k} \equiv f_{\bm k}(\bm r, t)$, the electron distribution function in this valley. It obeys the following kinetic equation 
\begin{equation}
\label{kin:gen}
\frac{\partial f_{\bm k}}{\partial t}+ {\bm v_{\bm k}\cdot}\bm \nabla_{\bm r} f_{\bm k} +\frac{1}{\hbar} \bm F\cdot \frac{\partial f_{\bm k}}{\partial \bm k} + Q_{imp}\{f_{\bm k}\} + Q_{ph}\{f_{\bm k}\}=0.
\end{equation}
Here $\bm F = e\bm E$ is the real force acting on the electrons in the presence of external electric field $\bm E$, $ Q_{imp}\{f_{\bm k}\}$ and $ Q_{ph}\{f_{\bm k}\}$ are the collision integrals for the electron-impurity and electron-phonon scattering, respectively. 
The forms of the collision integrals are specified below. 

In what follows we consider three situations:
\begin{enumerate}
\item[(i)]  VHE driven by the static electric field, in which case $\bm F = {\rm const}(t)$, time derivatives can be set to zero and the kinetic equation~\eqref{kin:gen} reduces to
\begin{equation}
\label{kin:F}
(\bm F\cdot \bm v_{\bm k}) f_0' + Q_{imp}\{f_{\bm k}\} + Q_{ph}\{f_{\bm k}\}=0, 
\end{equation}
where $f_0$ is the equilibrium distribution function with $f_0'=\partial f_0/\partial \varepsilon_k$ being the energy derivative of the equilibrium distribution.
\item[(ii)] VHE driven by the \emph{photon} drag effect at low frequencies $\omega$ of radiation, where $\omega\tau_p \ll 1$ with $\tau_p$ being the electron scattering time. In this case 
\begin{equation}
\label{photon:drag}
\bm F = \bm F_0 \exp{(\mathrm{i}\bm q\bm r-\mathrm i\omega t)} + {\rm c.c.},
\end{equation}
with $\bm q$ being the radiation wavevector, and the second-order response in ${\bm F_0}$ is calculated.
\item[(iii)] VHE due to the \emph{phonon} drag where $\bm F \equiv 0${, but}
in $Q_{ph}\{f_{\bm k}\}$ we take into account contributions proportional to the anisotropic part of the phonon distribution function~\cite{legurevich:drag,PhysRevB.100.045426}.
\end{enumerate}

The collision integral for electrons scattered by the static disorder contains the symmetric part responsible for the relaxation of the distribution function to the isotropic one and the asymmetric part responsible for the impurity skew scattering, ${Q_{imp}\{f_{\bm k}\} = Q_{imp}^{(s)}\{f_{\bm k}\} + Q_{imp}^{(as)}\{f_{\bm k}\}}$, where
\begin{subequations}
\label{collisions:imp}
\begin{align}
& Q_{imp}^{(s)}\{f_{\bm k}\} = \frac{f_{\bm k} - {\langle f_{\bm k}\rangle_{\varphi_{\bm k} }}}{\tau_{imp}}, \quad \frac{1}{\tau_{imp}} = \frac{2\pi}{\hbar} g U_c^2{n_i},\label{s:imp}\\
&  Q_{imp}^{(as)}\{f_{\bm k}\} = \sum_{\bm k'} W^{(as, imp)}_{\bm k'\bm k} f_{\bm k'}
.\label{as:imp}
\end{align}
Here ${\langle f_{\bm k}\rangle_{\varphi_{\bm k} }}= (2\pi)^{-1} \oint f_{\bm k} d\varphi_{\bm k}$, $\varphi_{\bm k}$ is the polar angle of the electron wavevector, $g=m/(2\pi\hbar^2)$ is the electron density of states, $U_{c,v}$ are assumed to be real, and $W^{(as, imp)}_{\bm k'\bm k}$ is the asymmetric (or skew) scattering probability for the electron-impurity scattering~\cite{dyakonov71a,sturmanBOOK}. Note that in collision integral responsible for the asymmetric scattering the out-scattering term ${\propto \sum_{\bm k'} W_{\bm k'\bm k}^{(as,imp)}f_{\bm k}}$ vanishes due to the general property ${W_{\bm k'\bm k}^{(as,imp)} = - W_{\bm k\bm k'}^{(as,imp)}}$~\cite{sturmanBOOK}. The skew-scattering contribution arises due to the asymmetric term $\propto [\bm k'\times \bm k]_z$ in Eq.~\eqref{Mcc} and results in the preferable scattering direction, Fig.~\ref{fig:intro}(a). We calculate it in the lowest non-vanishing order, namely, third order of perturbation theory:
\begin{multline}
\label{W:as:imp:3}
W_{\bm k' \bm k}^{as,imp} = \frac{2\pi}{\hbar} n_i \delta(\varepsilon_k -\varepsilon_{k'})\\
\times
\left|M_{\bm k'\bm k} +\sum_{\bm p} \frac{M_{\bm k'\bm p} M_{\bm p\bm k}}{\varepsilon_k - \varepsilon_{{p}} + \mathrm i 0} \right|^2
\end{multline}
 with the result~\cite{PhysRevB.75.045315,Sinitsyn_2007}:
\begin{equation}
\label{skew:imp}
W_{\bm k' \bm k}^{as,imp} = \xi S_{imp}[\bm k \times \bm k']_z  \delta(\varepsilon_{k}-\varepsilon_{k'}) ,
\end{equation}
where
\begin{equation}
\label{Simp}
S_{imp} = {{2\pi}U_v\over \tau_{imp}}.
\end{equation}
\end{subequations}

The situation with electron-phonon scattering is more complex. Generally, the collision integral contains four contributions:
\begin{subequations}
\begin{multline}
Q_{ph}\{f_{\bm k}\} =
 Q_{ph}^{(s)}\{f_{\bm k}\} + Q_{ph}^{(dr)}\{f_{\bm k}\}\\
 + Q_{ph}^{(as)}\{f_{\bm k}\}+ Q_{ph}^{(dr,as)}\{f_{\bm k}\}.\label{collisions:ph}
\end{multline}
 First one, $Q_{ph}^{(s)}\{f_{\bm k}\}$, describes quasi-elastic scattering resulting in the isotropization of the electron distribution function. It has the same form as the collision integral with impurities~\eqref{s:imp} with the phonon-induced scattering rate 
\begin{equation}
\label{s:ph}
\frac{1}{\tau_{ph}}=\frac{2\pi}{\hbar} g \Xi_c^2 \frac{k_B T}{\rho s^2}.
\end{equation}
Second contribution, $Q_{ph}^{(dr)}\{f_{\bm k}\}$, is responsible for the phonon drag effect. It describes the scattering of electrons by anisotropic part of the phonon distribution function, which is formed as a result of the lattice temperature gradient in the 2D  plane. In this situation, the phonon distribution function assumes the form
\begin{equation}
\label{ph:dist}
n_{\bm q} = \langle b_{\bm q}^\dag b_{\bm q}\rangle_{{ph}} = \frac{k_B T}{\hbar s q}\left(1 + {\bm q \over q} \cdot \bm e\right), \quad \bm e = -{s\tau_p^{ph}\over T} \bm \nabla T.
\end{equation}
Hereafter we omit subscript $\bm r$ in $\bm \nabla_{\bm r}$ for brevity and assume that the phonon and electron temperatures are the same; $\tau_p^{ph}$ is the phonon momentum relaxation time. Making use of  Fermi's golden rule we obtain~\cite{PhysRevB.100.045426}
\begin{equation}
\label{drag:ph}
 Q_{ph}^{(dr)}\{f_0\} = (\bm F_{\rm drag} \cdot \bm v_{\bm k}) f_0', 
\end{equation}
with
\begin{equation}
\label{drag:ph:force}
\bm F_{\rm drag} =-\frac{\tau_p^{ph}}{\rho \hbar} \left(\frac{m}{\hbar}\right)^2 \Xi_c^2 k_B \bm \nabla T.
\end{equation}
It follows from Eq.~\eqref{drag:ph} that the lattice temperature gradient produces an effective force $\bm F_{\rm drag}$ acting on the electrons. We stress that this force is not associated with any real potential energy and cannot be included in the Hamiltonian of the system, rather it appears in the kinetic equation from the collision integral. We also disregard the contribution to the current due to the Seebeck effect, i.e., the temperature gradient of the electron gas itself{, see Sec.~\ref{sec:discuss} for details}. Two remaining contributions to the collision integral describe the skew scattering of electrons by phonons:
\begin{align}
&Q_{ph}^{(as)}\{f_{\bm k}\} = \sum_{\bm k'} W^{(as, ph)}_{\bm k'\bm k} f_{\bm k'},\label{as:ph}\\
&Q_{ph}^{(dr,as)}\{f_{\bm k}\} = \sum_{\bm k'} W^{(as, dr)}_{\bm k'\bm k} f_0(\varepsilon_{k'}).\label{Q:dr:as}
\end{align}
The former contribution, similarly to Eq.~\eqref{as:imp} describes the elastic skew-scattering of electrons by phonons (this term does not contain the temperature gradient), while the latter one describes the asymmetric scattering in the course of the drag with $ W^{(as, dr)}_{\bm k'\bm k}  \propto \bm \nabla T$.
\end{subequations}

Importantly, the single-phonon processes described by Eq.~\eqref{v:def} do not contribute to the asymmetric scattering probabilities $ W^{(as, ph)}_{\bm k'\bm k}$, $W^{(as, dr)}_{\bm k'\bm k}$ even in the next-to-Born approximation. Indeed, the skew scattering results from the interference of the single and double scattering processes, but such interference is impossible because the number of phonons in the initial and final states is different for the two pathways. This is in stark contrast with the disorder scattering, Eq.~\eqref{v:dis}, where each impurity contributes to the skew scattering. Formally, the impurity potential in our model is not a Gaussian disorder, while the phonon-induced potential is Gaussian.
Thus, we need to take into account the two-phonon scattering processes~\cite{gy61,bs78} described by the Hamiltonian which we take in the simplest possible form [cf. Ref.~\cite{gantmakher87}]
\begin{multline}
\label{2ph}
{\tilde{V}_{c,v}} = \sum_{\bm q_1, \bm q_2} \tilde{\Xi}_{c,v} \frac{\hbar}{2\rho s} \sqrt{q_1q_2}\\
\times 
(b_{\bm q_1} {e}^{\mathrm i\bm q_1 \cdot \bm r - \mathrm i s q_1 t} - {\rm H.c})(b_{\bm q_2}{e}^{\mathrm i\bm q_2 \cdot \bm r - \mathrm i s q_2 t} -{\rm H.c}).
\end{multline}
Here $\tilde{\Xi}_{c,v}$ are the parameters describing the interaction of the conduction and valence band electrons with two phonons. The Hamiltonian~\eqref{2ph}  accounts for both the processes where two phonons are created or annihilated simultaneously and the Raman-like processes where the phonon is scattered by the electron. Following Ref.~\cite{bs78} we derive the contributions to the asymmetric scattering probabilities in the form [see Appendix~\ref{app:2ph} for details]
\begin{subequations}
\begin{equation}
\label{W:as:ph}
W_{\bm k' \bm k}^{(as,ph)} = \xi S_{ph} [\bm k \times \bm k']_z \delta(\varepsilon_k-\varepsilon_{k'}).
\end{equation}
Equation~\eqref{W:as:ph} is similar to Eq.~\eqref{skew:imp} with
\begin{equation}
\label{S:ph}
S_{ph} = -{4\pi\over\tau_{ph}} \left(\tilde{\Xi}_v+ 2{\Xi_v\over\Xi_c} \tilde{\Xi}_c \right){k_B T\over \rho s^2}.
\end{equation}
In derivation of $S_{ph}$ we  took into account the band mixing
both in the single and two-phonon scattering processes.
The expression for $W_{\bm k' \bm k}^{(as,dr)}$ is quite cumbersome, but in Eq.~\eqref{Q:dr:as} we need only the rate averaged over the angle of $\bm k'$. It reads
\begin{equation}
\label{W:as:dr}
\langle W_{\bm k' \bm k}^{(as,dr)}\rangle_{\varphi_{\bm k'}} =  -\frac{ms}{2\hbar} {\xi} S_{ph} [\bm e \times \bm k]_z\varepsilon_k\delta'(\varepsilon_k-\varepsilon_{k'}).
\end{equation}
\end{subequations}

Now we are able to solve the kinetic equation by iterations in the asymmetric scattering terms and calculate the valley Hall current due to the skew scattering mechanism for three effects discussed above. 
Hereafter we consider non-degenerate electrons to simplify calculations.\footnote{The results in the case of the static field and the photon drag remain the same for arbitrary degeneracy of the charge carriers provided that only impurity scattering is taken into account.}

\subsection{Skew scattering in the static electric field}\label{subsec:skew:F}

In the first order in the static electric field $\bm F$ the distribution function acquires a correction 
\begin{equation}
\label{corr:F}
\delta f_{\bm k}  = -(\bm F\cdot \bm v_{\bm k})f_0'\tau_p, \quad \frac{1}{\tau_p} =\frac{1}{\tau_{imp}}+ \frac{1}{\tau_{ph}}.
\end{equation}
This correction is substituted in the asymmetric collision integrals~\eqref{as:imp} and \eqref{as:ph} acting as a source of the skew contribution to the electron distribution responsible for the VHE:
\begin{equation}
\label{skew:F}
\frac{\delta f_{\bm k}^{(as)}}{\tau_p} + Q_{ph}^{(as)}\{ \delta f_{\bm  k}\} + Q_{imp}^{(as)}\{ \delta f_{\bm  k}\}=0.
\end{equation} 
Straightforward calculation of the valley Hall current
\begin{equation}
\label{VHC}
\bm j_{\rm VH} = e\sum_{\bm k} \bm v_{\bm k} \delta f_{\bm k}^{(as)} 
\end{equation}
yields
\begin{equation}
\label{jVHC:F}
\bm j_{\rm VH} = 2\xi (S_{ph} +S_{imp}) g {e\over {\hbar}}N\tau_p [\hat{\bm z}\times \bm F] {\bar \varepsilon \tau_p \over \hbar}.
\end{equation}
Here $N {=\sum\limits_{\bm k} f_0(\varepsilon_{\bm k})}$ is the electron density per valley and 
\[
\bar \varepsilon = {1\over N} \sum\limits_{\bm k} \varepsilon_{\bm k} f_0(\varepsilon_{\bm k}) 
\]
is the average electron energy.

\subsection{Skew scattering at the phonon drag}\label{subsec:skew:phonon}

It follows from Eq.~\eqref{collisions:ph} that there are two contributions to the skew scattering induced VHE under the phonon drag conditions. The first one is similar to that caused by the electric field and related to the rotation of the phonon drag-induced electric current due to the skew scattering. This contribution is given by Eq.~\eqref{jVHC:F} with the replacement $\bm F \to \bm F_{\rm drag}$. The second contribution is related to the skew scattering at the drag, Eq.~\eqref{Q:dr:as}. 
The corresponding correction to the electron distribution is found from the equation
\begin{equation}
\frac{\delta f_{\bm k}^{(as,dr)}}{\tau_p} + Q_{ph}^{(dr,as)} =0.
\end{equation}
Calculation shows that this contribution is given by
\[
\bm j_{\rm VH}' = -\xi S_{ph}g{e\over {\hbar}}N \tau_p [\hat{\bm z}\times \bm F_{\rm drag}]{\bar \varepsilon \tau_p \over \hbar}.
\]
Summing up these two contributions we arrive at the following expression for the valley Hall current under the phonon drag conditions due to the skew scattering
\begin{equation}
\label{jVHC:ph}
\bm j_{\rm VH}^{(ph)} = \xi (S_{ph} +2S_{imp}) g {e\over {\hbar}}N\tau_p[\hat{\bm z}\times \bm F_{\rm drag}]{\bar \varepsilon \tau_p \over \hbar}.
\end{equation}
We stress that the expression for the VHE under the phonon drag is not reduced to the one derived in the presence of a real force.

\subsection{Skew scattering at the photon drag}\label{subsec:skew:photon}

Let us now turn to the VHE under the photon drag conditions. Our aim is to illustrate the effect, this is why we consider the situation studied in Ref.~\cite{PhysRevLett.122.256801} and  termed by the authors ``Valley Acoustoelectric Effect''. This situation corresponds to the  oblique incidence of 
the $p$-polarized light,
so that the product $\bm F_0 \cdot \bm q_{\parallel}$ is nozero, where $\bm q_{\parallel}$ is the component of the light wavevector in the 2D
plane and $\bm F_0$ is the amplitude of the alternating force field acting on the electron, Eq.~\eqref{photon:drag}. 
In this geometry, the so-called $qE^2$ contribution dominates in the photon drag~\cite{perel73,Glazov2014101}.
Following Ref.~\cite{PhysRevLett.122.256801} we assume that the product $\omega\tau_p \ll 1$ and consider the leading in $(\omega\tau_p)^{-1}$ contribution to the photocurrent. 

At first, we calculate the \emph{dc} current in the absence of the skew scattering. Under the assumptions formulated above we obtain  in agreement with Ref.~\cite{PhysRevLett.122.256801}\footnote{The difference with Eq. (3) of Ref.~\cite{PhysRevLett.122.256801} (in the absence of screening) by the factor $4$ is related to the different definition of the electromagnetic wave amplitude, our Eq.~\eqref{photon:drag}.}
\begin{equation}
\label{dc:drag}
\bm j_{dc} = \frac{{N}e^2\tau_p^2 }{m^2\omega} (\bm q_{\parallel} \cdot \bm F_0)\bm F_{0,\parallel}^* + \text{c.c.} \equiv \frac{e{N}\tau_p }{m} \bm F_{\rm p, drag}. 
\end{equation}
Here we introduced the effective photon drag force
\begin{equation}
\label{photon:drag:force}
\bm F_{\rm p, drag} = \frac{e\tau_p}{m\omega}  (\bm q_{\parallel} \cdot \bm F_0)\bm F_{0,\parallel}^* + \text{c.c.}
\end{equation}
It is noteworthy that Eq.~\eqref{dc:drag} can be derived not only by solving the kinetic Eq.~\eqref{kin:gen} by iterations, but also from the simple arguments: In the first order in $\bm F_0$ the electron density is perturbed as $\delta {N} = \delta {N}_0 \exp{(\mathrm i \bm q_\parallel \bm r - \mathrm i \omega t)} +{\rm c.c.}$ in accordance with the continuity equation:
\begin{equation}
\label{cont}
\frac{\partial \delta {N}}{\partial t} + \bm\nabla\cdot \bm j_{ac}=0, \qquad \bm j_{ac} = \frac{e{N}\tau_p }{m} \bm F. 
\end{equation}
The \emph{dc} current~\eqref{dc:drag} arises as a result of the rectification as
\begin{equation}
\label{dc:drag:0}
\bm j_{dc} = \frac{e\tau_p}{m} \overline{\delta {N} \bm F},
\end{equation}
where the overline denotes the time-average.

Now we are able to calculate the valley Hall current. It arises as a result of the skew scattering of the anisotropically distributed electrons. Such anisotropic distribution is formed in the second order in $\bm F_0$ due to the photon drag. This contribution can be readily evaluated and has a form analogous to Eq.~\eqref{jVHC:F}
\begin{equation}
\label{jVHC:phot}
\bm j_{\rm VH}^{(phot)} =2\xi (S_{ph} +S_{imp})g {e\over {\hbar}}N\tau_p [\hat{\bm z} \times \bm F_{\rm p, drag}]{\bar \varepsilon \tau_p \over \hbar} .
\end{equation}
Note that accounting for the skew-scattering effect on the time dependent anisotropic distribution of electrons formed in the first order in the \emph{ac} field results in the parametrically smaller contribution to the VHE $\sim \bm j_{\rm VH}^{(phot)}\omega\tau_p$.

\subsection{Coherent skew scattering}\label{subsec:coh:skew}

There is a special skew-scattering contribution parametrically different from the above discussed ones. It arises from the coherent scattering by two closely spaced impurities or due to interference of the two single-phonon scattering processes, as discussed in Refs.~\cite{Ado_2015,PhysRevB.96.235148}. 
By contrast to the skew scattering considered above, it is present at Gaussian scattering potential as well~\cite{PhysRevB.75.045315}.
If the scattering is provided solely by the short-range disorder, Eq.~\eqref{v:dis}, and the electrons are driven by the static electric field, the corresponding expression for the asymmetric scattering rate is given by [cf. Eq.~\eqref{W:as:imp:3}]
\begin{multline}
\label{W:as:imp:2}
W_{\bm k' \bm k}^{as,2} = \frac{2\pi}{\hbar} {n_i^2\over 2} \delta(\varepsilon_k -\varepsilon_{k'})\\
\times
\left|\sum_{\bm p}\frac{M_{\bm k'\bm p}^{(2)} M_{\bm p\bm k}^{(1)}}{\varepsilon_k - \varepsilon_{p}+\mathrm i 0} + \sum_{\bm p'}\frac{M_{\bm k' {\bm p'}}^{(1)} M_{{\bm p'}\bm k}^{(2)}}{\varepsilon_k - \varepsilon_{{p'}}+\mathrm i 0} \right|^2.
\end{multline}
Here the superscripts $(1)$ and $(2)$ denote the impurities, and the factor $n_i^2/2$ accounts for the density of impurity pairs. After averaging the interference contribution over the impurity positions and picking out the asymmetric part, 
\begin{equation}
\propto \mathrm i\sum_{\bm p\bm p'} \frac{M_{\bm k \bm p'} M_{\bm p' \bm k'} M_{\bm k'\bm p} M_{\bm p\bm k}}{\varepsilon_k - \varepsilon_{{p'}}} \delta_{\bm p',\bm k + \bm k'-\bm p}\delta(\varepsilon_k-\varepsilon_p),
\end{equation}
we recover the result of the so-called X diagram in Refs.~\cite{Ado_2015,PhysRevB.96.235148} taken in the limit $\varepsilon_k \ll E_g$ (the so-called $\Psi$ diagram vanishes for the short-range scattering):
\begin{equation}
\label{j:X}
\bm j_{\rm VH}^\text{X} = -2{e\over {\hbar}}\xi \frac{U_v}{U_c} N[\hat{\bm z} \times \bm F].
\end{equation}
We stress that in this work we consider only $\xi$-linear contributions to the VHE, i.e., the current  proportional to $E_g^{-2}$. These are the lowest non-vanishing in $\bar \varepsilon /E_g$ contributions to the valley Hall current.
Accordingly, in Eq.~\eqref{W:as:imp:2}, we took into account only intraband scattering processes described by the matrix elements~\eqref{Mcc}, and ignored the contributions with the intermediate states in the valence band, Eq.~\eqref{Mcvvc}. The latter also provide the VH current, but it is parametrically smaller by a factor $\bar \varepsilon/E_g \ll 1$ as compared to Eq.~\eqref{j:X}, cf. Refs.~\cite{PhysRevB.75.045315,Ado_2015}.


Similar results with $\bm F\to \bm F_{\rm drag}, \bm F_{\rm p,drag}$ can be obtained at the drag conditions if the scattering asymmetry results from the impurities only. For the electron-phonon scattering such interference processes can be included as a renormalization of the two-phonon interaction constants $\tilde \Xi_{c,v}$ in Eq.~\eqref{2ph}.

\section{Anomalous contributions to VHE}
\label{sec:side}

We now switch to the contributions to the VHE beyond the kinetic equation approach. These contributions arise due to (i) the appearance of the anomalous contribution to the electron velocity in the electric field, Eq.~\eqref{v:anom}, caused by the spin-orbit interaction and (ii) the shifts of electronic wavepackets at the scattering events, so-called side-jump contribution. The variation of the electron coordinate at the scattering is given by~\cite{PhysRev.95.1154,belinicher82,Sturman2019,culcer2010}.
\begin{subequations}
\begin{multline}
\label{electron:shifts}
\bm R_{\bm k'\bm k}^{i} = - \frac{\Im\{M_{\bm k, \bm k'}^{i} (\bm \nabla_{\bm k} + \bm \nabla_{\bm k'})M_{\bm k', \bm k}^{i} \}}{|M_{\bm k' \bm k}|^2}  + \bm \Omega_{\bm k'} - \bm \Omega_{\bm k}
\\
=  \xi \frac{V_v(q)}{V_c(q)} [ (\bm k'-\bm k) \times \hat{\bm z}] + \xi[ (\bm k'-\bm k) \times \hat{\bm z}],
\end{multline}
where the scattering matrix element $M_{\bm k',\bm k}^{i}$ is given by Eq.~\eqref{Mcc} and the superscript $i$ distinguishes scattering mechansims (impurities or phonons). The second term in Eq.~\eqref{electron:shifts} does not depend on the scattering potential and related to the matrix elements $\bm \Omega_{\bm k}$ of the coordinate operator calculated on the Bloch functions~\eqref{e:Bloch}: $\bm \Omega_{\bm k}= i\left\langle u_{{c},\bm k} \left|\bm \nabla_{\bm k} \right|  u_{{c},\bm k}\right\rangle$. 
It is noteworthy that the shift of the electron coordinate in our case can be written as
\begin{equation}
\label{electron:shifts:111}
\bm R_{\bm k'\bm k} = \bm \varrho(\bm k') - \bm \varrho(\bm k),
\end{equation}
with $\bm \varrho(\bm k) = \xi[1+ V_v(q)/V_c(q)][\bm k \times \hat{\bm z}]$.
\end{subequations}
The side jump effect is illustrated in Fig.~\ref{fig:intro}(b) and the appearance of the anomalous velocity in Fig.~\ref{fig:intro}(c).

These anomalous velocity and side-jump effects largely cancel each other and call for special analysis~\cite{dyakonov_book}. Here we present the main results, and the detailed derivations are summarized in  Appendix~\ref{app:Keld}.

\subsection{Anomalous contributions in the static electric field}\label{subsec:anom:F}

The anomalous velocity contribution to the VHE can be readily calculated from Eq.~\eqref{v:anom} with the  result
\begin{equation}
\label{jVHC:va:F}
\bm j_{a} = -2{e\over {\hbar}} N\xi [\hat{\bm z}\times \bm F]. 
\end{equation}

First side-jump contribution to the valley Hall current can be found from the anisotropic field induced correction, $\delta f_{\bm k}$, Eq.~\eqref{corr:F} and reads
\begin{multline}
\label{sj:F:1}
\bm j_{sj}^{(1)} = e \frac{2\pi}{\hbar} \sum_{\bm k\bm k',i} \bm R^{i}_{\bm k'\bm k} |M^{i}_{\bm k',\bm k}|^2 \delta( \varepsilon_{k'} -  \varepsilon_{k}) \delta f_{\bm k} \\
= {e\over {\hbar}} \xi N\left(1+\frac{\tau_p}{\tau_{imp}}\frac{U_v}{U_c}+\frac{\tau_p}{\tau_{ph}}\frac{\Xi_v}{\Xi_c}\right) [\hat{\bm z} \times \bm F].
\end{multline}
We note that the scattering processes by impurities and phonons are not correlated, thus the resulting side-jump current contains three independent contributions: due to the electron coordinate operator $\bm\Omega_{\bm k}$ (first term, independent of the scattering mechanism), due to the impurity and phonon scattering (second and third terms, respectively).

Second side-jump contribution arises from the so-called anomalous distribution~\cite{PhysRevB.75.045315}
\begin{subequations}
\label{sj:F:adist}
\begin{equation}
\label{sj:F:adist:1}
\bm j_{sj}^{(2)} = e\sum_{\bm k} \bm v_{\bm k} \delta f^a_{\bm k},
\end{equation}
where the anomalous correction to the electron distribution function arises if we take into account the work of the force field $\bm F$ in the course of electron shift at the scattering:
\begin{multline}
\label{adist:2}
\frac{\delta f^a_{\bm k}}{\tau_p}
 = \frac{2\pi}{\hbar}\sum_{\bm k',i}  |M_{\bm k'\bm k}^i|^2  \\
  \times \delta(\varepsilon_{k'} -  \varepsilon_{k} - \bm R_{\bm k'\bm k}^i\cdot \bm F) [f_{0}(\varepsilon_{k'}) -  f_{0}(\varepsilon_k)].
\end{multline}
\end{subequations}
Decomposing the $\delta$-function in Eq.~\eqref{adist:2} up to the first order in $\bm F$ we obtain $\bm j_{sj}^{(2)} = \bm j_{sj}^{(1)}$.

As a result, the total anomalous current reads
\begin{multline}
\label{jVHE:anom:F}
\bm j_{\rm VH}^{(anom)}= \bm j_{a} + \bm j_{sj}^{(1)}+\bm j_{sj}^{(2)}\\
= 2{e\over {\hbar}} \xi N\left(\frac{\tau_p}{\tau_{imp}}\frac{U_v}{U_c}+\frac{\tau_p}{\tau_{ph}}\frac{\Xi_v}{\Xi_c}\right) [\hat{\bm z} \times \bm F].
\end{multline}
Notably, the resulting current is strongly sensitive to the scattering mechanisms, see  Sec.~\ref{sec:discuss} for details. {In the case of the electron scattering by a symmetric ($U_c = U_v$) short range 
disorder, Eq.~\eqref{jVHE:anom:F} is in agreement with previous works, Refs.~\cite{PhysRevB.75.045315,Ado_2015,PhysRevB.96.235148,Ando:2015aa}.}

\subsection{Anomalous contributions at the phonon drag}\label{subsec:anom:phonon}

The anomalous contributions to the VHE under the phonon drag conditions arise solely from the shifts of electronic wavevepackets at the scattering acts~\cite{belinicher82}. This is because there are no real static fields applied to the electrons and the anomalous velocity is absent, see Appendix~\ref{app:Keld} where this result is rigorously derived. Similarly to the skew scattering mechanism at the phonon drag there are two contributions to the side-jump current. First one, $\bm j_{sj}^{(ph,1)}$ similarly to the case of the external field,  results in the shifts calculated using anisotropic distribution function with the anisotropy induced by the phonon drag. It has the same form as Eq.~\eqref{sj:F:1} with the replacement $\bm F \to \bm F_{\rm drag}$. The second contribution arises due to the electronic  shifts in the course of drag. It is solely related to the electron-phonon scattering and takes the form
\begin{equation}
\label{sj:phonon:2}
\bm j_{sj}^{(ph,2)} =- {e\over {\hbar}} \xi N\left(1+  \frac{\Xi_v}{\Xi_c}\right)[\hat{\bm z} \times \bm F_{\rm drag}].
\end{equation}
The resulting current reads
\begin{multline}
\label{jVHE:anom:phonon}
\bm j_{\rm VH}^{(anom,ph)}= \bm j_{sj}^{(ph,1)}+ \bm j_{sj}^{(ph,2)}\\
= {e\over {\hbar}} \xi N { \frac{\tau_p}{\tau_{imp}}  \left(\frac{U_v}{U_c} - \frac{\Xi_v}{\Xi_c}\right) } [\hat{\bm z} \times \bm F_{\rm drag}].
\end{multline}
Importantly, this result does not reduce to VHE in the presence of a static electric field, Eq.~\eqref{jVHE:anom:F}.

\subsection{Anomalous contributions at the photon drag}\label{subsec:anom:photon}

Just as in Sec.~\ref{subsec:skew:photon} we consider the situation of low frequencies of radiation, $\omega\tau_p \ll 1$. Similarly to the analysis presented above, the side-jump contribution $\bm j_{sj}^{(phot,1)}$ arises from the shifts of electronic wavepackets at the impurity or phonon scattering. It has the form of Eq.~\eqref{sj:F:1} with the replacement $\bm F \to \bm F_{\rm p, drag}$, because this effect is not sensitive to particular mechanism leading to the anisotropic electron distribution.

There are two more contributions to the anomalous current specific to the photon drag mechanism. First of those was uncovered in Ref.~\cite{PhysRevLett.122.256801} can be associated with the anomalous velocity caused by the alternating field. Corresponding valley Hall current arises as a time-average [cf. Eq.~\eqref{dc:drag:0}]
\begin{equation}
\label{jVHC:va:photon}
\bm j_a^{(phot)} = e\overline{\bm v_a \delta N} = -2{e\over {\hbar}} N \xi[\hat{\bm z} \times \bm F_{\rm p, drag}].
\end{equation}
This expression is in agreement with Eq. (6) of Ref.~\cite{PhysRevLett.122.256801}. Second contribution can be associated with that due to the work of the \emph{ac} electric field and corresponding electronic shifts [cf. Eq.~\eqref{adist:2}]. Calculation of the time-average yields
\begin{multline}
\label{jVHC:sj:2:photon}
\bm j_{sj}^{(phot,2)} = \bm j_{sj}^{(phot,1)} \\
= {e\over {\hbar}} \xi N\left(1+\frac{\tau_p}{\tau_{imp}}\frac{U_v}{U_c}+\frac{\tau_p}{\tau_{ph}}\frac{\Xi_v}{\Xi_c}\right) [\hat{\bm z} \times \bm F_{\rm p, drag}].
\end{multline}
In this case the total anomalous current reads
\begin{multline}
\label{jVHE:anom:phot}
\bm j_{\rm VH}^{(anom,phot)}= \bm j_{a}^{(phot)} + \bm j_{sj}^{(phot,1)}+\bm j_{sj}^{(phot,2)}\\
= 2{e\over {\hbar}} \xi N\left(\frac{\tau_p}{\tau_{imp}}\frac{U_v}{U_c}+\frac{\tau_p}{\tau_{ph}}\frac{\Xi_v}{\Xi_c}\right) [\hat{\bm z} \times \bm F_{\rm p, drag}].
\end{multline}
It is noteworthy that for the low-frequency photon drag the anomalous contributions take the same form as the contributions induced by the real force field, Eq.~\eqref{jVHE:anom:F}.
We stress that the anomalous velocity contribution at the photon drag is totally compensated by a part of the side-jump contribution similarly to the case of the static field.

\section{Discussion}\label{sec:discuss}

\begin{table*}[t]
\caption{The valley Hall conductivity introduced as ${\bm j_\text{VH} = \sigma_\text{VH} [\hat{\bm z} \times {(\bm F/e)}]}$, where $\bm F$ is a force for static electric field, $\bm F_\text{drag}$ for phonon drag and $\bm F_\text{p, drag}$ for photon drag effect. Calculating the coherent skew contribution we assumed that the scattering is caused by both impurities and phonons but the scattering asymmetry is due to the impurities only. 
}
\label{tab}
 \begin{tabular}{c|c|c|c}
 \hline\hline
  Driving force  & {\bf skew} & {\bf coherent skew} & {\bf anomalous} \\ 
 \hline
 {\bf Electric field,} & \multirow{2}{*}{\hspace{0.2cm}$2{e^{{2}}\over {\hbar}} \xi N(S_{ph}+S_{imp})g\tau_p {\bar{\varepsilon}\tau_p\over\hbar}$\hspace{0.2cm}} & \multirow{2}{*}{\hspace{0.2cm}$-2{e^{{2}}\over {\hbar}} \xi N {\tau_p\over\tau_{imp}} {U_v\over U_c}$\hspace{0.2cm}} & \multirow{2}{*}{\hspace{0.2cm}$2{e^{{2}}\over {\hbar}} \xi N\left({\tau_p\over\tau_{imp}} {U_v\over U_c} + {\tau_p\over\tau_{ph}} {\Xi_v\over \Xi_c}\right)$\hspace{0.2cm}}\\ 
 {\bf photon drag} & &  & \\
 \hline 
 {\bf Phonon drag} &\hspace{0.2cm} ${e^{{2}}\over {\hbar}} \xi N(S_{ph}+2S_{imp})g\tau_p {\bar{\varepsilon}\tau_p\over\hbar}$\hspace{0.2cm} & \hspace{0.2cm}$-2{e^{{2}}\over {\hbar}} \xi N {\tau_p\over\tau_{imp}} {U_v\over U_c}$ \hspace{0.2cm} & \hspace{0.2cm} ${e^{{2}}\over {\hbar}} \xi N{\tau_p\over\tau_{imp}} \left({U_v\over U_c} -{\Xi_v\over \Xi_c}\right)$\hspace{0.2cm} \\
  \hline\hline
 \end{tabular}
\end{table*}

Above we have derived the contributions to the valley Hall current in three cases: (i) electrons are drifting in the static electric field, (ii) electrons are dragged by the electromagnetic wave, and (iii) electrons are dragged by the phonons.  
The results are summarized in Table~\ref{tab}.
Let us first compare the efficiency of the skew scattering and anomalous mechanisms of the VHE. We start with the situation where the dominant scattering mechanism is due to the static impurities where the disorder potential is given by Eq.~\eqref{v:dis}  and assume that the skew scattering is dominated by the standard third-order process. Taking into account~\eqref{Simp} and omitting numerical factors we obtain the following estimate for the skew scattering induced valley Hall current in agreement with Eqs.~\eqref{jVHC:F}, \eqref{jVHC:ph}, and \eqref{jVHC:phot}
\begin{equation}
\label{skew:VHE:est}
\mbox{skew:} \quad \bm j_{\rm VH}^{skew} \sim {e\over {\hbar}} {\xi} N [\hat{\bm z} \times \bm F] 
\frac{\bar \varepsilon \tau_{p}}{\hbar}
g U_v.
\end{equation}
Here $\bm F$ stands for any type of dragging force. For the anomalous contributions we obtain from Eqs.~\eqref{jVHE:anom:F}, \eqref{jVHE:anom:phonon}, and \eqref{jVHE:anom:phot}
\begin{equation}
\label{anom:VHE:est}
\mbox{anomalous:} \quad \bm j_{\rm VH}^{anom} \sim {e\over {\hbar}} {\xi} N [\hat{\bm z} \times \bm F] \frac{U_v}{U_c}.
\end{equation}
The ratio of the skew and anomalous currents is given by the product of the two factors
\[
\left|\frac{j_{\rm VH}^{skew}}{j_{\rm VH}^{anom}}\right| \sim  \frac{\bar \varepsilon \tau_{p}}{\hbar} {\times} g|U_c| {\sim \frac{\bar \varepsilon}{n_i |U_c|}}.
\]
The first factor (in the middle estimate) is assumed to be large in our approach, because it corresponds to freely propagating electrons with only rare scattering events. The second one controls the efficiency of scattering~\cite{ll3_eng}, and it is usually assumed to be small, $g|U_c| \ll 1$.\footnote{Strictly speaking, in two-dimensions for the short-range scattering $U_c$ should be replaced by the total scattering amplitude, which contains a logarithmic enhancement factor~\cite{ll3_eng}.} That is why, generally, the ratio of the skew and anomalous currents can be on the order of unity and both mechanisms should be taken into account. If the scattering is dominated by phonons, the parameter $g|U_c|$ is replaced by the efficiency of the two-phonon processes 
\[
\tilde \Xi {g} \frac{k_B T}{\rho s^2},
\]
which is less than unity at moderate temperatures where the acoustic phonon scattering is important~\cite{gantmakher87}. Thus, similarly to the case of the static disorder studied above, the skew-scattering and anomalous contributions can be comparable in the case of the electron-phonon interaction.

The situation becomes even more complex if we account for the coherent contribution to the skew scattering caused by the two-impurity processes, Sec.~\ref{subsec:coh:skew}. This coherent skew contribution, Eq.~\eqref{j:X} contains neither the large factor $\bar \varepsilon \tau_p/\hbar$, nor the small factor $g|U_c|$, and has the same form as the anomalous current, Eq.~\eqref{anom:VHE:est}. Importantly, if the scattering is solely caused by the static disorder, this contribution exactly compensates the anomalous current, {cf. Eq.~\eqref{jVHE:anom:F}} at $\tau_p=\tau_{imp}$. Note, that this cancellation is not general, it is absent if the scattering is anisotropic~\cite{PhysRevB.96.235148}.

Now let us discuss in more detail the anomalous contributions and, in particular, compensation of the anomalous velocity and side-jump contributions in the electric field and under the photon drag. Indeed, as it follows from Eqs.~\eqref{jVHE:anom:F} and \eqref{jVHE:anom:phonon} the anomalous velocity contribution was cancelled with the $\bm\Omega_{\bm k}$-contributions to the side-jump. 
Both contributions stem from the coordinate matrix element for Bloch electrons, $\bm \Omega_{\bm k} \propto \bm k$; the current is thus given by derivative of the coordinate operator, $\bm j_{\rm VH}^{\Omega} \propto d\bm \Omega_{\bm k}/dt {\propto d\bm k/dt}$. In the steady-state conditions, obviously, the time-derivative of the wavevector is zero, 
and this contribution to the current vanishes. Thus, these terms do not produce any \emph{dc} valley Hall current. 
The physical reason for this cancellation is clear~\cite{dyakonov_book,culcer2010}: 
The anomalous velocity due to the external force is compensated by the anomalous velocity at the scattering acts (generally speaking, due to the friction force), because at the steady-state the net force acting on the electron is zero.

This cancellation is also observed in the case of the phonon drag effect. It follows from Eq.~\eqref{jVHE:anom:phonon} that the valley Hall current depends on the scattering potentials and rates, while all contributions due to $\bm \Omega_{\bm k}$ vanish. Furthermore, if the impurity scattering is absent, $\tau_{imp} \to \infty$, then  $\bm j_{\rm VH}^{anom,ph}\equiv 0$. This is a general property of the electron shift current provided that $\bm R_{\bm k', \bm k}$ can be presented as a difference $\bm \varrho(\bm k') - \bm \varrho(\bm k)$ [cf. Eq.~\eqref{electron:shifts:111}], irrespectively of the scattering mechanism. Indeed, the steady-state kinetic equation for the distribution function~\eqref{kin:gen} can be recast in the simple form
\begin{subequations}
\begin{equation}
\label{kin:simple}
0= \sum_{\bm k'} \left(W_{\bm k',\bm k} f_{\bm k} - W_{\bm k,\bm k'} f_{\bm k'}\right),
\end{equation}
where the scattering rate $W_{\bm k',\bm k}$ includes both the momentum relaxation at the electron-phonon scattering and phonon drag. The side-jump current can be written by virtue of Eq.~\eqref{kin:simple} in the form [cf.~\cite{belinicher82}]
\begin{multline}
\label{sj:simple}
\bm j^{(anom,ph)}_{\rm VH} = \sum_{\bm k'\bm k} \left[\bm \varrho(\bm k') - \bm \varrho(\bm k)\right] W_{\bm k',\bm k} f_{\bm k}\\
=  \sum_{\bm k'\bm k} \bm \varrho(\bm k') \left(W_{\bm k',\bm k} f_{\bm k} - W_{\bm k,\bm k'} f_{\bm k'}\right) \equiv 0.
\end{multline}
\end{subequations}

In the presence of impurity scattering, this cancellation of side-jump currents at the phonon drag  is violated: Despite Eq.~\eqref{kin:simple} takes place with the total scattering probability $W_{\bm k',\bm k}$, the wavepacket shifts are caused by both impurities and phonons, and the corresponding currents are generally not compensated, see Eq.~\eqref{jVHE:anom:phonon}. Namely, the part of the side-jump current related to the $\bm \Omega_{\bm k}$ always vanishes by virtue of Eq.~\eqref{sj:simple}, and the  current results form the difference of shifts at the impurity  and phonon scattering 
\begin{equation}
\bm j_{\rm VH}^{(anom,ph)} = \frac{1}{2} \sum_{\bm k'\bm k}  (W^{imp}_{\bm k',\bm k} -W^{ph}_{\bm k',\bm k}) f_{\bm k} (\bm R^{imp}_{\bm k'\bm k}-\bm R^{ph}_{\bm k'\bm k}). \nonumber
\end{equation}

Above we considered the two-band model, Eq.~\eqref{H:el}. Our results can be readily generalized to the multiband description of the energy spectrum. Indeed, under the assumption that both the impurities and phonons do not mix the bands at the $\bm K_\pm$ valleys, other bands provide additive contributions to the VHE. As a result, in the expressions for the valley Hall current the products $\xi U_v,~\xi \Xi_v,~\xi \tilde{\Xi}_v$ are replaced, respectively, by the sums over the bands
\[
\sum_n \frac{\hbar^2 |p_{c,n}|^2}{{m_0^2}E_n^2} U_{n}, \quad \sum_n \frac{\hbar^2 |p_{c,n}|^2}{{m_0^2}E_n^2} \Xi_n, \quad \sum_n \frac{\hbar^2 |p_{c,n}|^2}{{m_0^2}E_n^2} \tilde \Xi_n,
\]
where $n$ enumerates the bands, $E_n$ is the corresponding band gap, and the conduction band is excluded from the summation.

It is instructive to present order-of-magnitude estimates of the VHE. To that end, we introduce the valley Hall conductivity $\sigma_{\rm VH}$ such that $\bm j_{\rm VH} = \sigma_{\rm VH} [\hat{\bm z} \times \bm E]$, where $\bm E = \bm F/e$ is the effective electric field associated with the force acting on the electrons, see Tab.~\ref{tab}. The anomalous contributions to the conductivity combine as $\sigma_{\rm VH}^{(anom)} \sim ({e^2}/{\hbar}) \xi N$. For TMDC MLs one has
$\xi = 10\ldots 100$~\AA$^2$ and for typical $N=10^{12}$~cm$^{-2}$ the product $\xi N \sim (1\ldots 10) \times 10^{-4}$. For the skew scattering mechanism $\sigma_{\rm VH}^{(skew)} \sim  \sigma_{\rm VH}^{(anom)}gU{\bar\varepsilon \tau}/{\hbar}$. Assuming $\bar\varepsilon \tau/\hbar \sim 10 \ldots 100$ and $gU \sim 0.1$ we get $\sigma_{\rm VH}^{(skew)}/\sigma_{\rm VH}^{anom} \sim 1\ldots 10$. Thus, depending on the parameters of the system the skew and anomalous contributions can be comparable. {It is noteworthy that the valley Hall conductivities for the case of external electric field and drag effects are similar in magnitude. The effective field acting on the charge carriers due to the phonon drag is typically somewhat smaller than in conventional transport experiments: For example in Ref.~\cite{Mak27062014} the applied electric field $E\sim 1000$~V$/$cm (0.5~V of the source-drain voltage at 5~$\mu$m sample), while at the lattice temperature gradient of $300$~K/$\mu$m which is possible for tightly focused optical excitation~\cite{PhysRevLett.120.207401,Perea-Causin:2019aa} the effective field is $\lesssim 250$~V$/$cm (estimated assuming $100$~\% phonon-electron momentum transfer). Thus, one can expect smaller, but measurable valley Hall current due to the phonon drag. While in electronic systems, the valley Hall effect can be more conveniently studied in standard transport experiments, in excitonic systems the phonon-induced fluxes can be dominant, because these particles are neutral, see Ref.~\cite{glazov2020skew} for details. } 

{In our calculations the intervalley scattering of electrons by defects or zone-edge phonons as well as the electron-electron collisions are excluded from consideration. These effects can reduce the magnitude of the valley Hall current similarly to the effects of spin-flip  and electron-electron scattering in the case of spin Hall effect in conventional semiconductors~\cite{gi:02,av03,0295-5075-87-3-37008}. However, 
for relevant materials and temperature the intravalley spin-conserving collisions included in the momentum relaxation time $\tau_p$ are the dominant ones which makes it possible to neglect the role of the intervalley interactions. }

In our work we have disregarded the intrinsic contribution to the VHE which arises in topologically nontrivial systems. This contribution, important if the  Fermi level lies in the topologically nontrivial band gap, is proportional to the Chern number characterizing the band gap. The effect is related to the current flowing along the one-dimensional edge channels formed in the topological structures and insensitive to a static disorder~\cite{Novokshenov}. However, the drag effects and, particularly, the role of inelasticity of electron-phonon interaction in the edge VHE requires separate analysis.

Here we focused on the situations where the electrons were dragged by the photons or phonons. In the latter case, the electrons are driven by the lattice temperature gradient. Naturally, the electrons can be also driven by the temperature gradient of the electron gas itself, i.e., by the Seebeck effect. The anomalous Hall effect is possible in this situation as well~\cite{xiao06,PhysRevB.90.075430,Adachi_2013}, see also~\cite{PhysRevB.87.245309}. {For estimates of the valley Hall effect induced by the electron temperature gradient in the non-degenerate gas one can use the results in Tab.~\ref{tab} derived for the electric field and using the drag force in the form $\bm F_{\rm Seebeck} = (\mu/T) \bm \nabla  T$ with $\mu$ being the chemical potential~\cite{PhysRev.135.A1505,PhysRevB.100.045426,PhysRevB.101.155204,PhysRevB.101.195126,glazov2020skew}. The Seebeck and phonon drag contributions to the valley Hall current can be separated, e.g., via their temperature dependence, which enters $\tau_p^{ph}$ in Eq.~\eqref{drag:ph:force} [cf. Ref.~\cite{Mahan2014}] and also in the case where the lattice and electron temperatures gradients differ, e.g., due to an efficient electron-electron interaction.
The detailed analysis of the interplay of the skew-scattering, side-jump and anomalous velocity contributions in this case is still an open problem beyond the scope of this paper.}

It is interesting to note that the inverse effect  (IVHE) is also possible in 2D Dirac materials. 
For the VHE driven by electric field, in the inverse  effect, the
 valley current flowing in the sample is converted in an electric current in the transverse direction.
For the VHE in the phonon drag conditions, the IVHE is an occurrence of a lattice temperature gradient in the direction transverse to the valley current: {$\bm \nabla T_{\rm latt} \propto [\hat{\bm z} \times \bm j_\text{VH}]$}.
The microscopic mechanisms of IVHE are the same as in VHE: skew-scattering, anomalous velocity and side-jumps. It is not obvious if the Onsager relation takes place for the VHE and IVHE conductivities under phonon drag because there are no terms $\propto \bm F_{\rm drag}$ 
in the free energy. Therefore the IVHE microscopic calculation is a problem for separate study.

\section{Summary}\label{sec:concl}

Here we presented the microscopic theory of the valley Hall effect, i.e., generation of the valley current transversal to the external drag force, in 2D Dirac materials. 
The key result
is that VHE current is not universal and depends strongly on the scattering mechanisms in both conduction and valence bands. 
Our main focus was on the situation where the charge carriers are dragged by the non-equilibrium flux of phonons or by electromagnetic wave. We took into account all relevant contributions to the effect: the skew-scattering, the side-jump, and the anomalous velocity. Two latter contributions can largely compensate each other, and the contribution from skew scattering is dominant in many cases. Importantly, the valley Hall current depends on the source of the drag force.

\acknowledgments

The financial support of the Russian Science Foundation (Project~17-12-01265) is acknowledged.
The work of L.E.G. was supported by the Foundation for the Advancement of Theoretical Physics and Mathematics ``BASIS''. 

%

\appendix

\section{Phonon skew scattering rates}
\label{app:2ph}

\subsection{Elastic scattering}

At the phonon scattering, with account for both one- and two-phonon interactions, Eq.~\eqref{Mcc} holds where the matrix elements  $V_{c,v}(q)$ are equal to the sums of one- and two-phonon matrix elements $V_{c,v} + \tilde{V}_{c,v}$ given by Eqs.~\eqref{v:def} and~\eqref{2ph}.
In the third order of the perturbation theory we obtain, similarly to the impurity scattering, 
\begin{align}
W_{\bm k' \bm k}^{(as,ph)} & = {(2\pi)^2\over \hbar} \delta(\varepsilon_{k}-\varepsilon_{k'}) 
\nonumber \\ & \times
\sum_{\bm k_1} \text{Im}(M_{\bm k \bm k'}M_{\bm k' \bm k_1}M_{\bm k_1 \bm k})\delta(\varepsilon_{k}-\varepsilon_{k_1}),
\end{align}
which yields
\begin{align}
& W_{\bm k' \bm k}^{(as,ph)} =  \xi{(2\pi)^2\over \hbar} \delta(\varepsilon_{k}-\varepsilon_{k'}) g 
 \\ & \times
\left< (V_c + \tilde{V}_c)^2(V_v + \tilde{V}_v)[\bm k \times \bm k' + \bm k' \times \bm k_1 +  \bm k_1 \times \bm k]_z\right>
,\nonumber
\end{align}
where averaging is performed over both the angle $\varphi_{\bm k_1}$ and the phonon states.
In the linear order in both the valence-band constants and two-phonon interaction strength we have
\begin{align}
\label{W_skew}
W_{\bm k' \bm k}^{(as,ph)} = &  \xi{(2\pi)^2\over \hbar} \delta(\varepsilon_{k}-\varepsilon_{k'}) g 
 \\ \times & 
\biggl< (V_c^2\tilde{V}_v + \tilde{V}_cV_cV_v + V_c\tilde{V}_cV_v) \nonumber
\\ & \times
[\bm k \times \bm k' + \bm k' \times \bm k_1 +  \bm k_1 \times \bm k]_z \biggr>
. \nonumber
\end{align}
The first term in round brackets has the following form:
\begin{widetext}
\begin{multline}
\left<V_c^2\tilde{V}_v \right>_{ph} = -\Xi_c^2\tilde{\Xi}_v\left({ \hbar\over2\rho s} \right)^2 
\Biggl< \sum_{\bm q,\bm q'}\sqrt{q}(b_{\bm q}\delta_{\bm q, \bm k' - \bm k} - b^\dag_{\bm q}\delta_{\bm q, \bm k - \bm k'}) 
\sqrt{q'}(b_{\bm q'}\delta_{\bm q', \bm k_1 - \bm k'} - b^\dag_{\bm q'}\delta_{\bm q', \bm k' - \bm k_1}) \\
\times
\sum_{\bm q_1, \bm q_2} \sqrt{q_1q_2}
(b_{\bm q_1}b_{\bm q_2}\delta_{\bm q_1+\bm q_2, \bm k - \bm k_1}
- b_{\bm q_1}b_{\bm q_2}^\dag\delta_{\bm q_1-\bm q_2, \bm k - \bm k_1}
- b_{\bm q_1}^\dag b_{\bm q_2}\delta_{\bm q_2-\bm q_1, \bm k - \bm k_1}
+ b_{\bm q_1}^\dag b_{\bm q_2}^\dag \delta_{\bm q_1+\bm q_2, \bm k_1 - \bm k}) \Biggr>_{ph}.
\end{multline}
\end{widetext}
Averaging over phonon states yields:
\begin{align}
\label{ph_skew1}
\left<V_c^2\tilde{V}_v \right>_{ph} =  & -8\Xi_c^2\tilde{\Xi}_v\left({ \hbar\over2\rho s} \right)^2 \\ 
& \times |\bm k - \bm k'| |\bm k_1 - \bm k'|  \:  \bar{n}_{|\bm k' - \bm k|}  \bar{n}_{|\bm k' - \bm k_1|}, \nonumber
\end{align}
where $\bar{n}_q=(n_{\bm q}+n_{-\bm q})/2$ is the even part of the phonon distribution ($n_{\bm q}=\langle b^\dag_{\bm q}b_{\bm q}\rangle_{ph}$).
Deriving this expression, we took into account that the terms with $\delta_{\bm k\bm k_1}$ coming from the product $\left<V_c^2\right>_{ph}\langle\tilde{V}_v\rangle_{ph}$ do not contribute to the skew scattering probability because ${\bm k \times \bm k' + \bm k' \times \bm k_1 +  \bm k_1 \times \bm k=0}$ at $\bm k = \bm k_1$.
%
Assuming equilibrium phonon distribution: $\bar{n}_q=k_\text{B}T/(\hbar s q)$, we obtain
\begin{equation}
\left<V_c^2\tilde{V}_v \right>_{ph} = - 2\Xi_c^2\tilde{\Xi}_v \left({k_\text{B}T\over \rho s^2}\right)^2.
\end{equation}

Similar calculations of two remaining terms yield:
\begin{equation}
\left<\tilde{V}_cV_cV_v\right>_{ph} =\left<V_c\tilde{V}_cV_v\right>_{ph} =- 2\tilde{\Xi}_c\Xi_c\Xi_v \left({k_\text{B}T\over \rho s^2}\right)^2.
\end{equation}
Averaging Eq.~\eqref{W_skew} over directions of $\bm k_1$ we obtain 
Eq.~\eqref{W:as:ph}: $W_{\bm k' \bm k}^{(as,ph)} = \xi S_{ph} [\bm k \times \bm k']_z \delta(\varepsilon_k-\varepsilon_{k'})$ with $S_{ph}$ given by Eq.~\eqref{S:ph} of the main text.

\subsection{Inelastic scattering}

Scattering of symmetrically distributed electrons, e.g. equilibrium electron gas, by drifting phonons results in the valley Hall effect. However, it is not described by the anisotropic skew scattering probability derived in the previous subsection. In order to quantify this effect, we take into account inelasticity of the electron-phonon interaction in the first order in the ratio of the characteristic phonon and electron energies $\hbar s q_\text{char}/k_\text{B}T \sim \sqrt{ms^2/k_\text{B}T} \ll 1$ with  $q_{\rm char} \sim \sqrt{mk_B T/\hbar^2}$.

\begin{widetext}
The valley-dependent scattering probability with account for the phonon energies has the following form
\begin{align}
W_{\bm k' \bm k}^{(as,dr)} = {(2\pi)^2\over \hbar} \sum_{\bm k_1} & \sum_{\mu, \nu = \pm 1}\text{Im}(M_{\bm k \bm k'}M_{\bm k' \bm k_1}M_{\bm k_1 \bm k})
\delta(\varepsilon_{k}-\varepsilon_{k'}+\nu\hbar s |\bm k'-\bm k|)
\delta(\varepsilon_{k'}-\varepsilon_{k_1}+\mu\hbar s |\bm k_1-\bm k'|). 
\end{align}

This results in the following modification of 
Eq.~\eqref{W_skew}:
\begin{align}
\label{sk_inel_1}
W_{\bm k' \bm k}^{(as,dr)}  =  - 2\xi{(2\pi)^2\over \hbar} 
 \left({ \hbar\over2\rho s} \right)^2   
 & \sum_{\nu, \mu=\pm 1, \bm k_1} [\bm k \times \bm k' + \bm k' \times \bm k_1 +  \bm k_1 \times \bm k]_z
\\
\times
\biggl[&\Xi_c^2\tilde{\Xi}_v m_{\nu(\bm k' - \bm k)}\delta(\varepsilon_{k}-\varepsilon_{k'}+\nu\hbar s |\bm k'-\bm k|)
 m_{\mu(\bm k_1 - \bm k')} \delta(\varepsilon_{k'}-\varepsilon_{k_1}+\mu\hbar s |\bm k_1-\bm k'|)
\nonumber\\
+ &\tilde{\Xi}_c\Xi_c\Xi_v m_{\nu(\bm k_1 - \bm k')}\delta(\varepsilon_{k'}-\varepsilon_{k_1}+\nu\hbar s |\bm k'-\bm k_1|)
 m_{\mu(\bm k_1 - \bm k)}\delta(\varepsilon_{k}-\varepsilon_{k_1}+\mu\hbar s |\bm k-\bm k_1|)
\nonumber\\
+ &\Xi_c\tilde{\Xi}_c\Xi_v m_{\nu(\bm k' - \bm k)}\delta(\varepsilon_{k}-\varepsilon_{k'}+\nu\hbar s |\bm k'-\bm k|)
 m_{\mu(\bm k - \bm k_1)}\delta(\varepsilon_{k_1}-\varepsilon_{k}+\mu\hbar s |\bm k_1-\bm k|)
\biggr] .\nonumber
\end{align}
Here $m_{\bm q}=qn_{\bm q}$.

We take the following general nonequilibrium phonon distribution 
\begin{equation}
n_{\bm q} = \bar{n}_q \left(1 + {c_{\bm q}\over q} \right), \qquad c_{-\bm q} = -c_{\bm q},
\end{equation}
where $\bar{n}_q=k_\text{B}T/(\hbar s q)$ is the Planck function. 
Then we have in the lowest order in electron-phonon scattering inelasticity
\begin{align}
W_{\bm k' \bm k}^{(as,dr)}= -4\Xi_c\xi{(2\pi)^2\over \hbar} 
 \left({k_\text{B}T\over 2 \rho s^2} \right)^2 \hbar s \sum_{\bm k_1} & [\bm k \times \bm k' + \bm k' \times \bm k_1 +  \bm k_1 \times \bm k]_z
\nonumber\\
\times
\biggl\{  & c_{\bm k' - \bm k}\delta'(\varepsilon_{k}-\varepsilon_{k'})  \left[\Xi_c\tilde{\Xi}_v \delta(\varepsilon_{k'}-\varepsilon_{k_1}) + \tilde{\Xi}_c\Xi_v \delta(\varepsilon_{k_1}-\varepsilon_{k})\right] 
\nonumber\\
 +
& 
c_{\bm k_1 - \bm k'}\delta'(\varepsilon_{k'}-\varepsilon_{k_1}) \left[\Xi_c\tilde{\Xi}_v \delta(\varepsilon_{k}-\varepsilon_{k'})+ \tilde{\Xi}_c\Xi_v\delta(\varepsilon_{k}-\varepsilon_{k_1}) \right]
\nonumber\\
+ &c_{\bm k_1 - \bm k}\delta'(\varepsilon_{k}-\varepsilon_{k_1})\tilde{\Xi}_c\Xi_v \left[\delta(\varepsilon_{k'}-\varepsilon_{k_1})+ \delta(\varepsilon_{k}-\varepsilon_{k'}) \right]
\biggr\} .\nonumber
\end{align}

Hereafter we assume the following asymmetry of the phonon distribution:
\begin{equation}
\label{c_q}
c_{\bm q} = {\bm q} \cdot \bm e
\qquad \text{i.e.} \qquad n_{\bm q} = {k_\text{B}T \over \hbar s q} \left(1 + {\bm q \over q} \cdot \bm e\right),
\end{equation}
where $\bm e$ is  the in-plane vector describing the direction of the phonon flux [cf. Eq.~\eqref{ph:dist}]. 
Then summation over $\bm k_1$ and integration over directions of $\bm k'$ yields
\begin{equation}
\left< W_{\bm k' \bm k}^{(as,dr)} \right>_{\varphi_{\bm k'}}= \xi{(2\pi)^2\over \hbar} 
 \left({k_\text{B}T\over \rho s^2} \right)^2 g{ms\over\hbar}(\bm e \times \bm k)_z
 \Xi_c(\Xi_c\tilde{\Xi}_v + 2\tilde{\Xi}_c\Xi_v)
  \left[ \varepsilon_{k'}\delta'(\varepsilon_{k}-\varepsilon_{k'})
-\delta(\varepsilon_{k'}-\varepsilon_{k})
\right].
\end{equation}
This expression is equivalent to Eq.~\eqref{W:as:dr} of the main text. 
\end{widetext}

Note that the same results can be derived using the Keldysh technique following the theory of Ref.~\cite{bs78}.

\section{Calculation of VHE in the Keldysh technique}\label{app:Keld}

In this section we present the main steps of the derivations of the anomalous contributions (side-jump and anomalous velocity) to VHE within the Keldysh diagram technique. This technique is appropriate for calculating quantum contributions to the transport properties of the non-equilibrium systems, i.e., for the electrons under the phonon or photon drag conditions and also makes it possible to evaluate the VHE in the presence of static electric field.

We introduce the Keldysh Greens functions $G^{\alpha\beta}_{n\bm k}(\epsilon)$, where the superscripts $\alpha,\beta=\pm$ indicate the line of the Keldysh contour, $n=c$ or $v$ indicates the band, $\epsilon$ is the energy variable. In the equilibrium conditions the functions read
\begin{align}
&G^{--}_{n\bm k}(\epsilon) = \frac{1-f_0(\varepsilon_{n,k})}{\epsilon - \varepsilon_{n,k} + \mathrm i\hbar/(2\tau_p)}+ \frac{f_0(\varepsilon_{n,k})}{\epsilon - \varepsilon_{n,k} - \mathrm i\hbar/(2\tau_p)},\nonumber \\
&G^{-+}_{n\bm k}(\epsilon) = \frac{f_0(\varepsilon_{n,k})}{\epsilon - \varepsilon_{n,k} - \mathrm i\hbar/(2\tau_p)} - \frac{f_0(\varepsilon_{n,k})}{\epsilon - \varepsilon_{n,k} + \mathrm i\hbar/(2\tau_p)},\nonumber 
\\
& G^{+-}_{n\bm k}(\epsilon) = \frac{1-f_0(\varepsilon_{n,k})}{\epsilon - \varepsilon_{n,k} + \mathrm i\hbar/(2\tau_p)} - \frac{1-f_0(\varepsilon_{n,k})}{\epsilon - \varepsilon_{n,k} - \mathrm i\hbar/(2\tau_p)},\nonumber \\
& G^{++}_{n\bm k}(\epsilon) = - [G^{--}_{n\bm k}(\epsilon)]^*.\label{Keldysh}
\end{align}
We note that the average value of any physical quantity $A$ via the corresponding
matrix elements $A_{nn'}({\bm k})$
\begin{equation}
\label{Keldysh:aver}
\langle A\rangle = \sum_{n,n', \bm k} \int_{-\infty}^\infty \frac{d\epsilon}{2\pi \mathrm i} A_{nn'}({\bm k}) G^{-+}_{n'n,\bm k}(\epsilon).
\end{equation}
Here $G^{-+}_{n'n,\bm k}$ are the interband ($n\neq n'$) and intraband ($n=n'$) matrix elements of the Keldysh $G^{-+}$ Greens function. The former appear due to the band mixing by the scattering, Eq.~\eqref{Mcvvc} and the external electric field.
In what follows we consider non-degenerate electrons. Thus, it is sufficient to disregard quadratic and higher-order contributions in $f_0(\varepsilon_{n,k})$ while evaluating the diagrams. As a result, we retain $f_0(\varepsilon_{n,k})$ in the $G_{n,k}^{-+}(\varepsilon)$ only, while put $f_0=0$ in the remaining Greens functions. Note that in calculation of the valence band Greens function it is sufficient to neglect the dispersion putting $\varepsilon_{v,k} = - E_g$ and disregard the occupancy of the valence band. We calculate the contributions to the VHE proportional to the electron density.
Below we put $\hbar =1$.

\subsection{VHE in the presence of a static electric field}\label{sec:app:F}

\begin{figure}[h]
 \includegraphics[width=\linewidth]{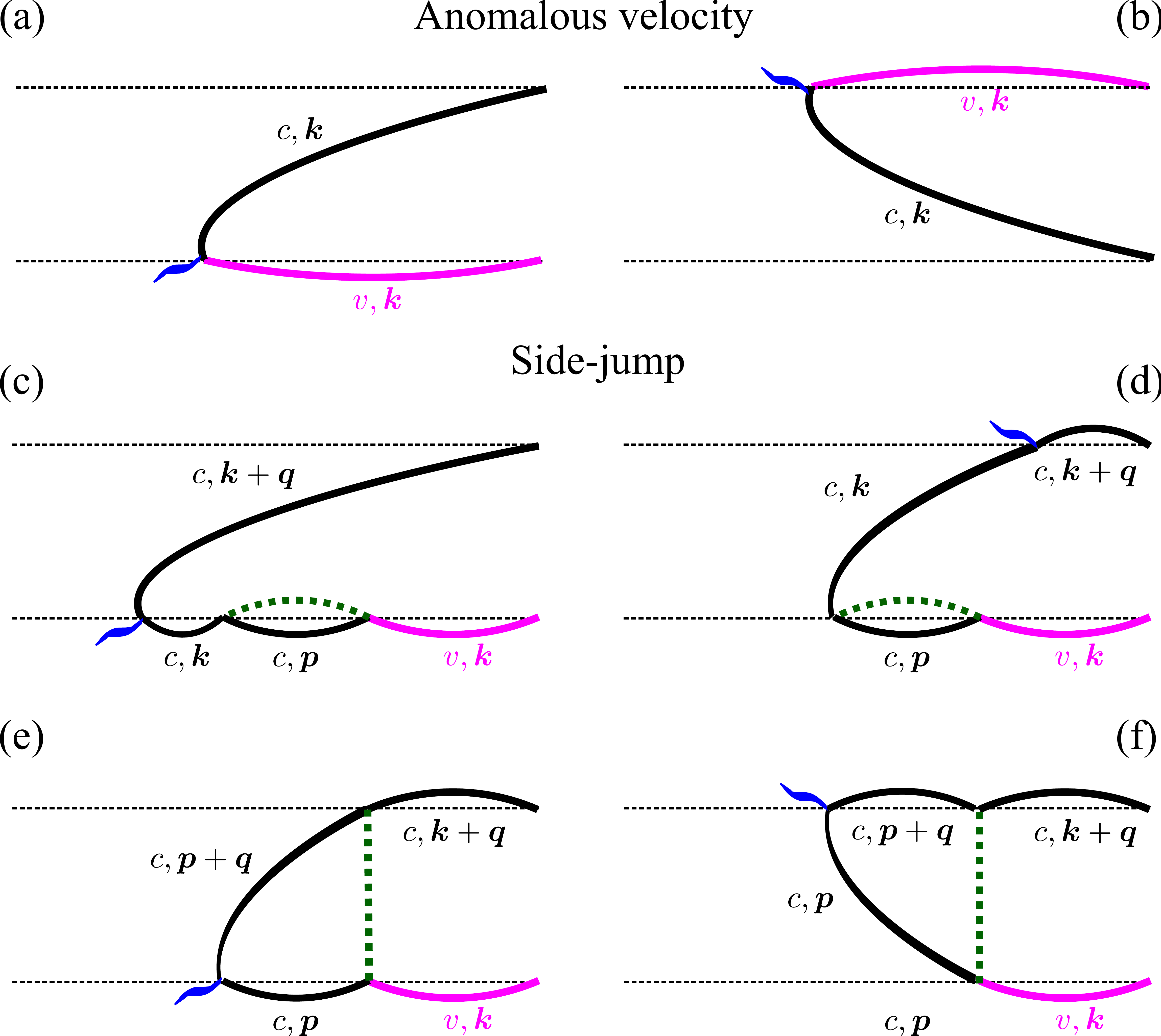}
\caption{Diagrams describing the anomalous velocity contribution (a,b) and side-jump contributions (c-f) in the electric field. Wavy line describes coupling with the electric field, dotted line describes scattering by the phonons or impurities. For the side-jump diagrams (c-f) their counterparts flipped along the horizontal axis are not shown.
}\label{app:F}
\end{figure}

Figures~\ref{app:F}(a) and~(b) show the diagrams which provide the anomalous velocity contribution to the electric-field induced VHE.
Let the \emph{ac} electric field $\bm E \parallel x$ oscillates in time at low frequency $\omega {\ll 1/\tau_p}$ which will be taken equal to zero in the end of calculation. 
The correction to the inter-band Green function shown in Fig.~\ref{app:F}(a) is given by
\begin{equation}
\label{dG:AVEfield}
\mathrm i \delta G^{-+}_{cv,\bm k}(\epsilon) = \mathrm i G^{-+}_{c,\bm k}(\epsilon) \mathrm i
\left(-\frac{e p_{cv}}{c m_0} A_x\right) \mathrm i G_{v,\bm k}^{++}(\epsilon - \omega).
\end{equation}
The diagram shown in Fig.~\ref{app:F}(b) is similar to the one in panel (a) and can be obtained by its flipping along the horizontal axis. It  provides the  contribution obtained from that in panel (a) by complex conjugation and the replacement ${\omega \to -\omega}$. 
Taking into account that $A_x = c E_x/(\mathrm i \omega)$ 
we see that the vertex of interaction with the field on the upper/lower contour equals to $\pm \gamma eE_x/\omega$.
Extracting {the} $\omega$-independent contribution 
by using ${G_{v,\bm k}^{++}(\epsilon - \omega)\approx -1/E_g + (\epsilon - \omega)/E_g^2}$
we obtain 
\begin{equation}
\label{dG:AVEfield:1}
 \mathrm i \delta G^{-+}_{cv,\bm k}(\epsilon) = - \mathrm i {\gamma}
\frac{e E_x}{E_g^2} 2\pi f_0(\varepsilon_k) \delta (\epsilon - \varepsilon_k).
\end{equation}
Here we took into account that the contribution $\propto 1/\omega$ is irrelevant and vanishes with account for both Fig.~\ref{app:F}(a) and (b).
According to Eq.~\eqref{Keldysh:aver}, the $y$-component of the interband VHE current reads
\begin{equation}
\label{i:anom:Efield}
j_{a,y} =  e\sum_{\bm k} \int \frac{d\epsilon}{2\pi\mathrm i}  \: \mathrm i {\gamma} \:
\delta G_{cv,\bm k}^{-+} = - e 
{\gamma^2\over E_g^2}
N E_x,
\end{equation}
%
or
\begin{equation}
\label{i:anom:Efield:fin}
\bm j_{a+b} = - 2e\xi  N[\hat{\bm z} \times (e\bm E)],
\end{equation}
in agreement with Eq.~\eqref{jVHC:va:F} (we recall that $\bm F = e\bm E$).

 Diagrams shown in Fig.~\ref{app:F}(c,d) describe the side-jump contribution related to the Bloch coordinate shift 
in the course of the scattering.
For calculation of these diagrams it is convenient to use the stationary gauge $\bm A=\bm 0$, ${\bm E = -\bm \nabla_{\bm r}\phi=-\mathrm i \bm q \phi_{\bm q} }$.
The vertex of interaction with the field on the upper/lower contour equals to $\mp \mathrm i e\phi_{\bm q}$.
The sum of diagrams shown in Fig.~\ref{app:F}(c,d) yields the following correction to the inter-band Green function
\begin{align}
&\mathrm i \delta G^{-+}_{cv,\bm k}(\epsilon) =\mathrm i e\phi_{\bm q} 
\sum_{\bm p} (-M_{\bm k,\bm p}M^{cv}_{\bm p,\bm k}) \mathrm i G^{++}_{c,\bm p}(\epsilon)  \mathrm i G_{v,\bm k}^{++}(\epsilon) \nonumber \\
&\times [\mathrm i G^{-+}_{c,\bm k+\bm q}(\epsilon)\mathrm i G^{++}_{c,\bm k}(\epsilon)
- \mathrm i G^{--}_{c,\bm k+\bm q}(\epsilon)\mathrm i G^{-+}_{c,\bm k}(\epsilon)].
\end{align}
Hereafter the scattering matrix elements $M_{\bm k, \bm p} = V_c(q)$ [Eq.~\eqref{Mcc} without $\bm k$-dependent terms], and $M^{cv}_{\bm p,\bm k}$ is given by Eq.~\eqref{Mcvvc}.
Taking $G_{v,\bm k}^{++}(\epsilon)=-1/E_g$ we have for the contribution to VHE current similarly to Eq.~\eqref{i:anom:Efield} after integration over $\epsilon$:
\begin{align}
&j_{c+d,y}^{(sj)} =  
e^2 \mathrm i \phi_{\bm q} \tau_p \xi \\ 
&\times \sum_{\bm k, \bm p}
\left( k_-  - {V_v\over V_c}p_-\right) {|M_{\bm k,\bm p}|^2 [f_0(\varepsilon_{\bm k+\bm q})-f_0(\varepsilon_k)]\over \mathrm i(\varepsilon_k-\varepsilon_p-\mathrm i/\tau_p)}+ {\rm c.c.} \nonumber
\end{align}
Here we neglected $\bm q$ in the denominator and ${\rm c.c.}$ stands for the contribution of the diagrams obtained by flipping the Fig.~\ref{app:F}(c,d) along the horizontal line, see discussion of the anomalous velocity contribution above.

Similarly, the sum of diagrams Fig.~\ref{app:F}(e,f)
yields the following correction to the Green function
\begin{align}
&\mathrm i \delta G^{-+}_{cv,\bm k}(\epsilon) =\mathrm i e\phi_{\bm q}
\sum_{\bm p} M_{\bm k,\bm p}M^{cv}_{\bm p,\bm k} \mathrm i G^{--}_{c,\bm k+\bm q}(\epsilon)  \mathrm i G_{v,\bm k}^{++}(\epsilon) \nonumber \\
&\times [\mathrm i G^{-+}_{c,\bm p+\bm q}(\epsilon)\mathrm i G^{++}_{c,\bm p}(\epsilon)
- \mathrm i G^{--}_{c,\bm p+\bm q}(\epsilon)\mathrm i G^{-+}_{c,\bm p}(\epsilon)].
\end{align}
Then the contribution to the VHE current density with account for the flipped diagrams as well yields
\begin{align}
&j_{e+f,y}^{(sj)} = 
-e^2 \mathrm i \phi_{\bm q} \tau_p \xi \\ 
&\times \sum_{\bm k, \bm p}
\left({V_v\over V_c}p_-  - k_- \right) {|M_{\bm k,\bm p}|^2 [f_0(\varepsilon_{\bm p+\bm q})-f_0(\varepsilon_p)]\over \mathrm i(\varepsilon_p-\varepsilon_k+\mathrm i/\tau_p)}+ \text{c.c.} \nonumber
\end{align}
Changing here notations $\bm p \leftrightarrow \bm k$, we obtain for the sum of the diagrams~(c) -- (f)
\begin{equation}
\bm j_{c-f}^{(sj)}=e \sum_{\bm k, \bm p} \bm R_{\bm p,\bm k} 2\pi |M_{\bm k,\bm p}|^2 \delta(\varepsilon_k-\varepsilon_p) \delta f_{\bm k},
\end{equation}
where we made expansion to the linear order in $\bm q$ and introduced $\delta f_{\bm k}$ according to Eq.~\eqref{corr:F} and the shift $\bm R_{\bm p,\bm k}$ according to Eq.~\eqref{electron:shifts}. This expression coincides with Eq.~\eqref{sj:F:1} of the main text. Therefore
a sum of the contributions (c) -- (f) yields:
\begin{equation}
\bm j_{c-f}^{(sj)} =  e\xi  N \left(1+\frac{\tau_p}{\tau_{imp}}\frac{U_v}{U_c}+\frac{\tau_p}{\tau_{ph}}\frac{\Xi_v}{\Xi_c}\right)[\hat{\bm z} \times (e\bm E)].
\end{equation}

\begin{figure}[h]
 \includegraphics[width=\linewidth]{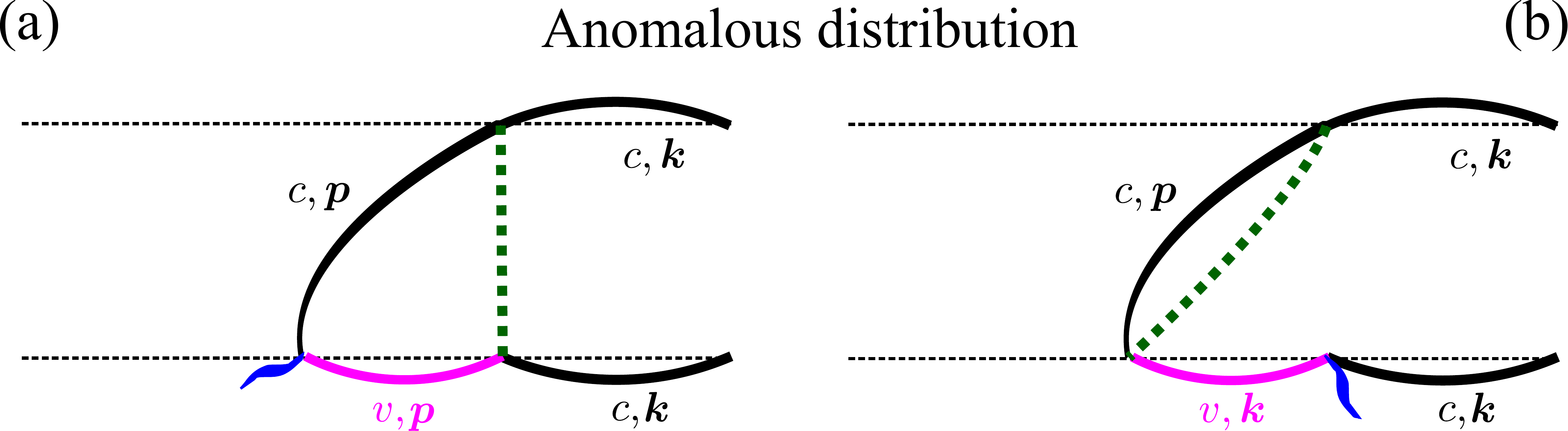}
\caption{Diagrams describing the side-jump contributions in the electric field due to the anomalous distribution of electrons. The flipped counterparts are not shown.
}\label{app:F:adist}
\end{figure}

The contributions due to the anomalous distribution of the electrons, Eqs.~\eqref{sj:F:adist}, are illustrated in Fig.~\ref{app:F:adist} (as before, the flipped diagrams are not shown). 
The correction to the conduction-band Green function depicted by the diagram (a) 
is given by
\begin{align}
\mathrm i \delta G^{-+}_{c,\bm k}(\epsilon) = &-{eE_x\gamma\over\omega}
\mathrm i G^{--}_{c,\bm k}(\epsilon+\omega)\mathrm i G^{++}_{c,\bm k}(\epsilon)  \nonumber \\
&\times \sum_{\bm p} M_{\bm k,\bm p}M^{vc}_{\bm p,\bm k}  \mathrm i G^{-+}_{c,\bm p}(\epsilon+\omega)\mathrm i G^{++}_{v,\bm p}.
\end{align}
Substituting the scattering matrix elements Eqs.~\eqref{Mcc},~\eqref{Mcvvc}, we get
\begin{align}
\delta G^{-+}_{c,\bm k}(\epsilon) =\mathrm i\xi {eE_x\over\omega} & G^{--}_{c,\bm k}(\epsilon+\omega)G^{++}_{c,\bm k}(\epsilon) \\
 \times  \sum_{\bm p}|M_{\bm k,\bm p}|^2& \left({V_v\over V_c}k_+-p_+\right)G^{-+}_{c,\bm p}(\epsilon+\omega). \nonumber
\end{align}
In the lowest order in $\omega$ we have after integration over $\epsilon$ for the contribution of the diagram Fig.~\ref{app:F:adist}(a) and its flipped counterpart:
\begin{align}
&j_{a,y}^{(adist)}= e\xi eE_x\tau_p
\\
&\times \sum_{\bm k,\bm p}|M_{\bm k,\bm p}|^2 v_{\bm k,y}\left({V_v\over V_c}k_y-p_y\right) f_0(\varepsilon_p)[-2\pi\delta'(\varepsilon_k-\varepsilon_p)]. \nonumber
\end{align}
This yields
\begin{equation}
\bm j_{a}^{(adist)} =  e\xi  N 
\left(\frac{\tau_p}{\tau_{imp}}\frac{U_v}{U_c}+\frac{\tau_p}{\tau_{ph}}\frac{\Xi_v}{\Xi_c}\right)[\hat{\bm z} \times (e\bm E)].
\end{equation}

The correction to the Green function depicted by the diagram Fig.~\ref{app:F:adist}(b) is given by
\begin{align}
\mathrm i \delta G^{-+}_{c,\bm k}(\epsilon) = &-{eE_x\gamma\over\omega}
\mathrm i G^{--}_{c,\bm k}(\epsilon+\omega)\mathrm i G^{++}_{c,\bm k}(\epsilon)   \\
&\times \sum_{\bm p} M_{\bm k,\bm p} M^{cv}_{\bm p,\bm k}  \mathrm i G^{-+}_{c,\bm p}(\epsilon+\omega)\mathrm i G^{++}_{v,\bm k}.\nonumber
\end{align}
Again after integration over $\epsilon$ we get for the contribution of the diagram Fig.~\ref{app:F:adist}(b) and its flipped counterpart
\begin{align}
&j_{b,y}^{(adist)}= e\xi eE_x\tau_p
\\
&\times \sum_{\bm k,\bm p}|M_{\bm k,\bm p}|^2 v_{\bm k,y}\left(k_y-{V_v\over V_c}p_y\right) f_0(\varepsilon_p)[-2\pi\delta'(\varepsilon_k-\varepsilon_p)]. \nonumber
\end{align}
Calculation yields
\begin{equation}
\bm j_{b}^{(adist)} =  e\xi  N [\hat{\bm z} \times (e\bm E)].
\end{equation}

One can also see that the sum of all four contributions depicted in Fig.~\ref{app:F:adist} has the form ($\bm F=e\bm E$)
\begin{equation}
\bm j_{a+b}^{(adist)}=2\pi\tau_p e\bm F\sum_{\bm k,\bm p}|M_{\bm k,\bm p}|^2 \bm R_{\bm p,\bm k}f_0(\varepsilon_p)\delta'(\varepsilon_p-\varepsilon_k),
\end{equation}
coinciding with $\bm j_{sj}^{(2)}$ given by Eq.~\eqref{sj:F:adist:1}:
\begin{equation}
\label{app:sj:2}
\bm j_{sj}^{(2)} 
= e\xi N\left(1+\frac{\tau_p}{\tau_{imp}}\frac{U_v}{U_c}+\frac{\tau_p}{\tau_{ph}}\frac{\Xi_v}{\Xi_c}\right) [\hat{\bm z} \times \bm F].
\end{equation}

Note that the diagrams presented in Figs.~\ref{app:F} and \ref{app:F:adist} are fully analogous to the diagrams presented in Ref.~\cite{PhysRevB.75.045315}.

\subsection{VHE caused by the phonon drag}\label{sec:app:phonon}

\begin{figure}[h]
 \includegraphics[width=\linewidth]{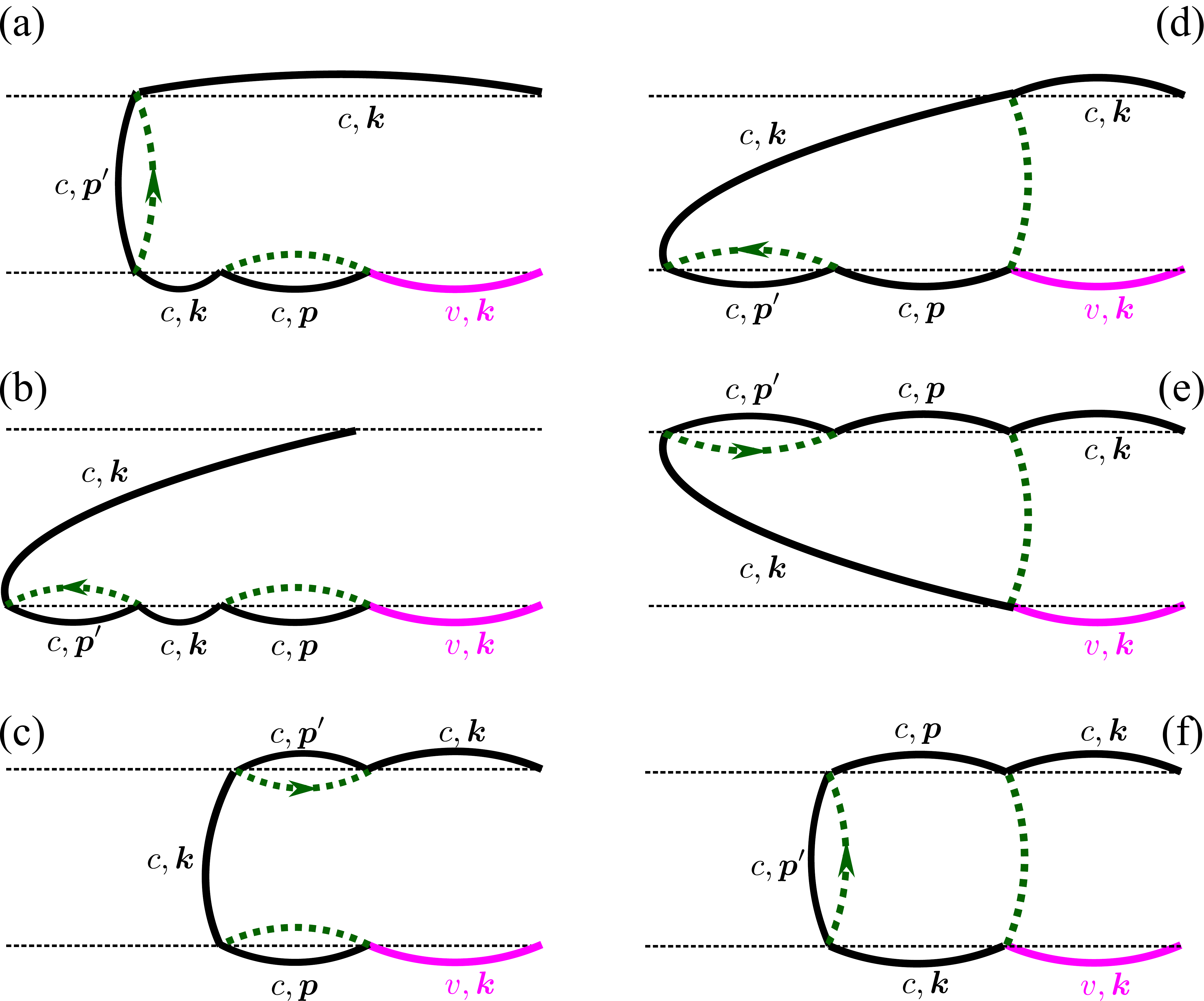}
\caption{Diagrams describing side-jump contributions  to the VHE under the phonon drag conditions. Left column (a-c) shows the contributions proportional to the Bloch electron coordinate shifts, the right column (d-f) shows the contributions related to the phase of the matrix element [first term in Eq.~\eqref{electron:shifts}]. Dotted line with arrow denotes the anisotropic part of the phonons Greens function which results in the electron drag. The flipped counterparts of these diagrams provide complex conjugate contribuions and are not shown.}\label{app:phonon}
\end{figure}

In the case of the phonon drag one has to take into account the self-energies which include anisotropic part of the phonon distribution function. Corresponding phonon lines are denoted by the dotted lines with arrow, Fig.~\ref{app:phonon}. This self-energy part plays a role of the electron-field interaction version in the case of the static field, Sec.~\ref{app:Keld}\ref{app:F}.
Accordingly, the correction to the inter-band Green function shown in Figs.~\ref{app:phonon}(a)-(c) is conveniently expressed as follows
\begin{equation}
\mathrm i \delta G^{-+}_{cv,\bm k}(\epsilon) = -M^{cc}_{\bm k,\bm p}M^{cv}_{\bm p,\bm k}  \mathrm i  G^{++}_{c,\bm p}(\epsilon) \mathrm i  G^{++}_{v,\bm k}(\epsilon)\mathrm i \delta G_1,
\end{equation}
where (omitting some terms nullifying after integration over $\epsilon$)
\begin{equation}
\delta G_1 = \delta f_{\bm k} \left({1\over \epsilon -\varepsilon_k-\mathrm i/2\tau_p} - {1\over \epsilon -\varepsilon_k+\mathrm i/2\tau_p}\right)
\end{equation}
with $\delta f_{\bm k}$ being the anisotropic electron distribution caused by the phonon drag:
\begin{equation}
\delta f_{\bm k} = -\tau_p Q_{ph}^{(dr)}\{f_0(\varepsilon_k)\} = f_0'(\varepsilon_k)\tau_p (\bm v_{\bm k}\cdot \bm F_\text{drag}).
\end{equation}
Calculation of the corresponding contribution to VHE current yields
\begin{equation}
j_{a+b+c,y}^{(sj)} = e \xi \sum_{\bm k, \bm p} \left( k_x  - {V_v\over V_c}p_x\right) 2\pi |M_{\bm k,\bm p}|^2 \delta(\varepsilon_k-\varepsilon_p) \delta f_{\bm k}.
\end{equation}

Evaluation of the diagrams shown in Figs.~\ref{app:phonon}(d)-(f) yields
\begin{equation}
j_{d+e+f,y}^{(sj)} = e \xi \sum_{\bm k, \bm p} \left({V_v\over V_c}p_x  - k_x \right) 2\pi |M_{\bm k,\bm p}|^2 \delta(\varepsilon_k-\varepsilon_p) \delta f_{\bm p}.
\end{equation}
Changing here notations $\bm p \leftrightarrow \bm k$, we obtain the sum of the diagrams~(a)-(f) in the form
\begin{equation}
\bm j_{a-f}^{(sj)}=e \sum_{\bm k, \bm p} \bm R_{\bm p,\bm k} 2\pi |M_{\bm k,\bm p}|^2 \delta(\varepsilon_k-\varepsilon_p) \delta f_{\bm k},
\end{equation}
coinciding with $\bm j_{sj}^{(ph,1)}$ from Sec.~\ref{subsec:anom:phonon}:
\begin{multline}
\label{sj:ph:1}
\bm j_{sj}^{(ph,1)} = e\xi N\left(1+\frac{\tau_p}{\tau_{imp}}\frac{U_v}{U_c}+\frac{\tau_p}{\tau_{ph}}\frac{\Xi_v}{\Xi_c}\right) [\hat{\bm z} \times \bm F_{\rm drag}].
\end{multline}

\begin{figure}[h]
 \includegraphics[width=0.9\linewidth]{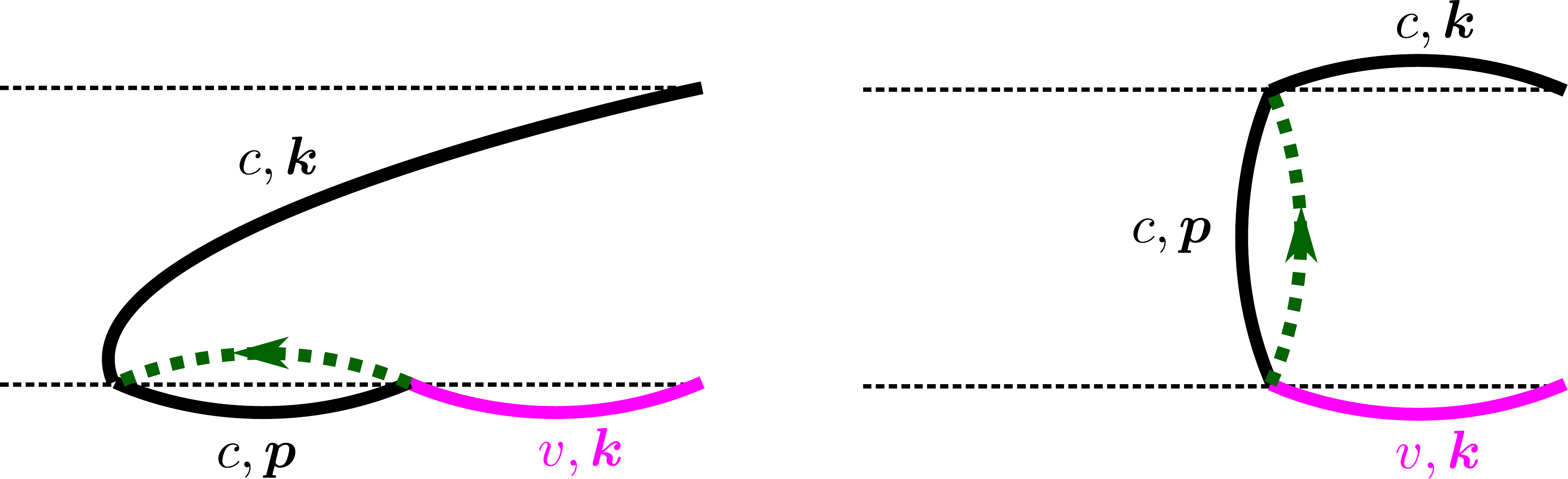}
\caption{Diagrams describing side-jump contributions  to the VHE arising in the course of the phonon drag, which are analogues of the anomalous distribution contribution in the electric field, Fig.~\ref{app:F:adist}. The flipped counterparts of these diagrams provide complex conjugate contribuions and are not shown.}\label{app:phonon:adist}
\end{figure}

Additional contribution to the valley Hall current arises, similarly to that of the anomalous distribution in the presence of the electric field,  due to electron wavepackets shifts in the course of drag. It is described by two diagrams (and their flipped counterparts) in Fig.~\ref{app:phonon:adist}. 

Evaluation of these diagrams yields 
\begin{align}
j_y =&e\xi\sum_{\bm k, \bm p} \left({\Xi_v\over \Xi_c}p_x  - k_x \right) \\ 
&\times \left[Q_{ph}^{(dr)}\{f_0(\varepsilon_k)\}-Q_{ph}^{(dr)}\{f_0(\varepsilon_p)\}\right]  \nonumber \\
=&-e\xi N \left(1 + {\Xi_v\over \Xi_c}\right)[\hat{\bm z} \times \bm F_{\rm drag}]_y.\nonumber
\end{align}
This result coincides with $\bm j_{sj}^{(ph,2)}$ given by Eq.~\eqref{sj:phonon:2}.

Note that the diagrams with the crossing phonon lines have additional smallness $\hbar/(\bar\varepsilon\tau_p)$. The diagrams with the valence band Greens function in the other positions have extra smallness $1/E_g$ as compared with the presented ones.

\subsection{VHE caused by the photon drag}\label{sec:app:photon}

\begin{figure}[h]
 \includegraphics[width=\linewidth]{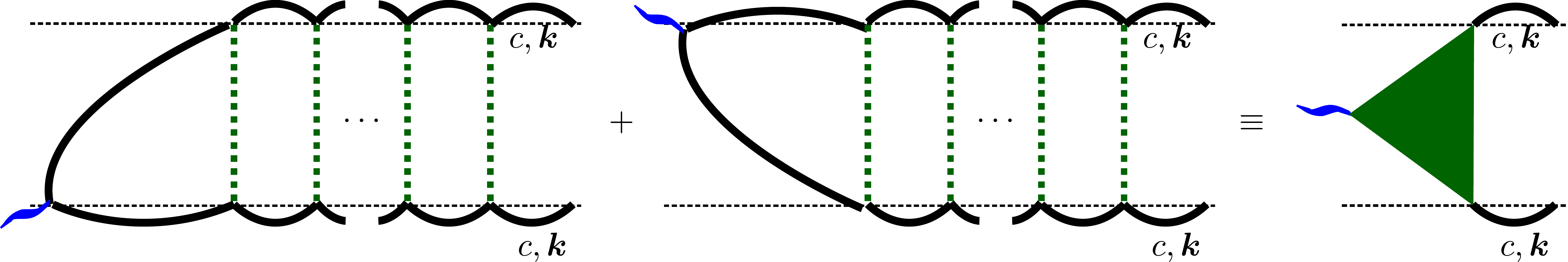}
\caption{Sum of the ladder diagrams relevant for the photon drag at $\omega\tau_p \ll 1$.}\label{app:ladder}
\end{figure}

In order to calculate the VHE under the photon drag conditions we, as in the main text, consider the limit $\omega\tau_p \ll 1$. It is instructive to calculate the first order in the electric field correction to the Greens function $G^{-+}$ depicted in Fig.~\ref{app:ladder} with the result {(note that the summation of the ladder diagrams is required here because the corresponding correction to the Greens function does not depend on the direction of the electron wavevector, thus the ladder does not vanish in for the short-range scattering and gives $\tau_p/\mathrm i\omega$ factor)}:
\begin{equation}
\label{qE:1}
\mathrm i G_1^{-+} = \frac{\tau_p}{2 \mathrm i \omega} v_k^2 (\bm q \bm F_0)f_0' \left(\frac{1}{\epsilon - \varepsilon_{ k} - \frac{ \mathrm i }{2\tau_p}} - \frac{1}{\epsilon - \varepsilon_{ k} + \frac{ \mathrm i}{2\tau_p}}\right).
\end{equation}

\begin{figure}[h!]
 \includegraphics[width=\linewidth]{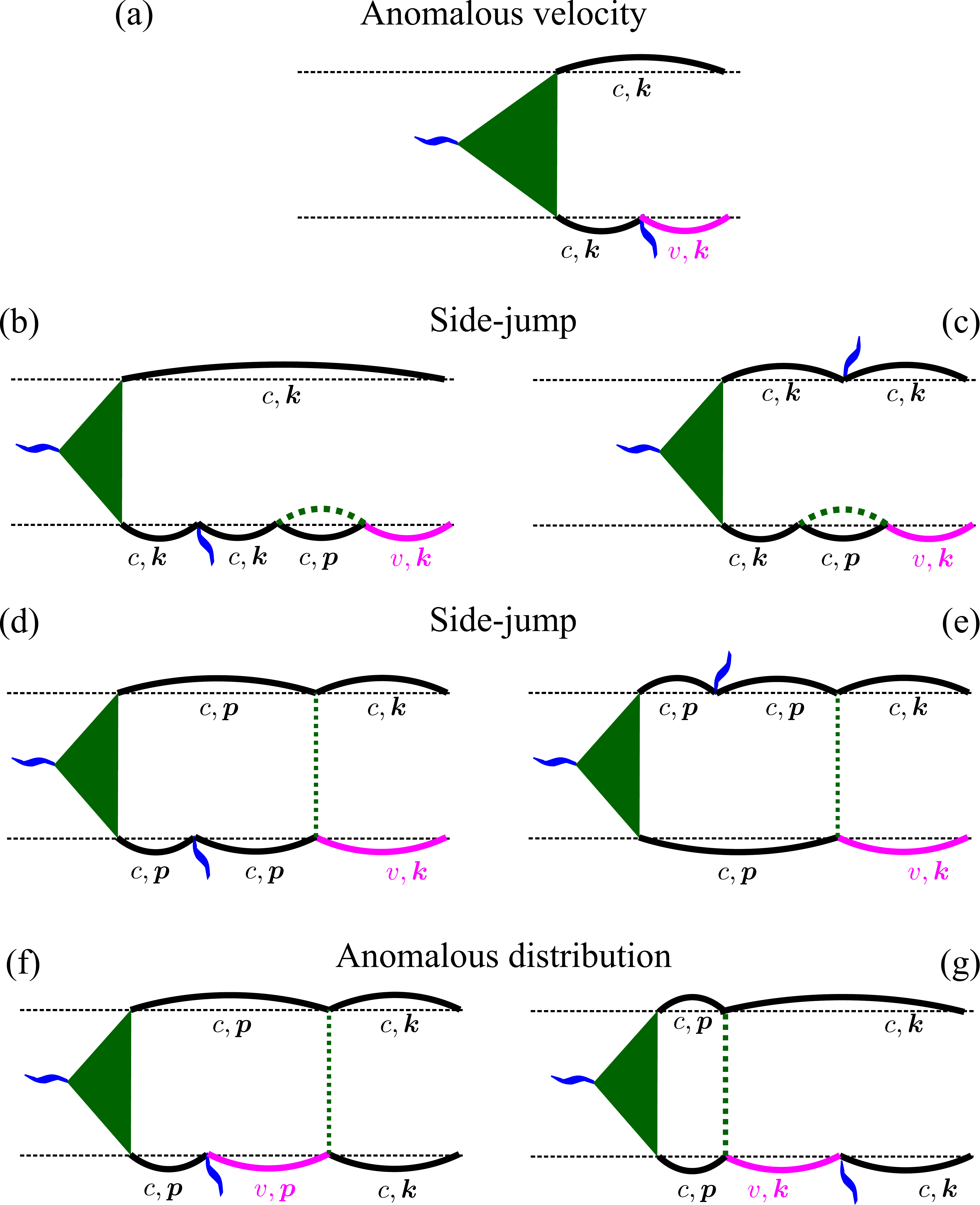}
\caption{Diagrams describing the anomalous velocity contribution (a), side-jump contributions (b-e), and anomalous distribution (f,g) at the photon drag conditions. Filled triangle denotes sum of the ladder diagrams in Fig.~\ref{app:ladder}. The flipped counterparts of these diagrams doubling the results are not shown.}\label{app:photon}
\end{figure}

Figure~\ref{app:photon} shows relevant diagrams for the VHE under the photon drag conditions. 
These diagrams are equivalent to those presented in Figs.~\ref{app:F} and~\ref{app:F:adist} with the only change $G_{c,\bm k}^{-+} \to G_1^{-+}$. Therefore we immediately obtain the results for the photon drag induced VHE from those in the static field by the substitution
\begin{equation}
f_0(\varepsilon_k)\bm F \to -\frac{\tau_p}{2 \mathrm \omega} v_k^2 (\bm q \bm F_0)f_0' \bm F_0^* + \text{c.c.} = -\varepsilon_k f_0'(\varepsilon_k)\bm F_{\rm p, drag}.
\end{equation}
Moreover, all results of Sec.~\ref{app:Keld}\ref{app:F} are valid after the substitution $\bm F \to \bm F_{\rm p, drag}$ because $\sum\limits_{\bm k}\varepsilon_k [-f_0'(\varepsilon_k)] {=\sum\limits_{\bm k}f_0(\varepsilon_k)=N}$.
In particular, 
diagram Fig.~\ref{app:photon}(a) with its flipped counterpart gives Eq.~\eqref{jVHC:va:photon}, Figs.~\ref{app:photon}(b-e) describe the side-jump contribution, 
\begin{multline}
\label{jVHC:sj:1:photon}
 \bm j_{sj}^{(phot,1)} \\
= e\xi N\left(1+\frac{\tau_p}{\tau_{imp}}\frac{U_v}{U_c}+\frac{\tau_p}{\tau_{ph}}\frac{\Xi_v}{\Xi_c}\right) [\hat{\bm z} \times \bm F_{\rm p, drag}],
\end{multline}
and the diagrams in Fig.~\ref{app:photon}(f,g) give the anomalous distribution contribution, Eq.~\eqref{jVHC:sj:2:photon}.

Note that the summation of the ladder at the  second field vertex in the diagrams Fig.~\ref{app:photon} is not needed because this vertex represents the anisotropic correction to the Greens function and the ladder diagrams vanish at the short-range scattering.

\newpage

\[
\mbox{ }
\]
\newpage

%

\end{document}